\documentclass[fleqn,usenatbib]{mnras}

\usepackage{newtxtext,newtxmath}

\usepackage[T1]{fontenc}

\DeclareRobustCommand{\VAN}[3]{#2}
\let\VANthebibliography\thebibliography
\def\thebibliography{\DeclareRobustCommand{\VAN}[3]{##3}\VANthebibliography}

\usepackage{graphicx}	
\usepackage{amsmath}	
\usepackage{hhline}

\newcommand{\triB}[1]{\langle #1 \rangle}
\newcommand{\absB}[1]{\left| #1 \right|}
\newcommand{\anglB}[1]{\left(#1\right)}
\newcommand{\squB}[1]{\left[#1\right]}

\newcommand{\updown}[3]{{#1^{#2}}_{#3}}
\newcommand{\mdot}{\dot{M}}
\newcommand{\mdotname}{mass flow rate }
\newcommand{\mdotin}{\dot{M}_{in}}
\newcommand{\mdotout}{\dot{M}_{out}}
\newcommand{\mdotun}{\dot{M}_{un}}
\newcommand{\mad}{\phi}
\newcommand{\flux}{\Phi}
\newcommand{\rsp}{r}
\newcommand{\rcyl}{\mathcal{R}}
\newcommand{\rescale}{\lambda}
\newcommand{\height}{\mathcal{H}}
\newcommand{\covdev}{\mathcal{D}}
\graphicspath{{pictures/}}

\title[Winds and jets in resistive GRMHD simulations]{Double Blandford: Resistive GRMHD simulations of accreting black holes simultaneously launch winds and jets}

\author[J. Speicher et al.]{J. Speicher, $^{1}$\thanks{Corresponding author: j.i.speicher@uva.nl}, O. Porth$^{1}$ and N. Degenaar$^{1}$ \\
     $^{1}$Anton Pannekoek Institute for Astronomy, University of Amsterdam, Science Park 904, NL-1090 GE Amsterdam, The Netherlands   }

\date{Accepted XXX. Received YYY; in original form ZZZ}

\pubyear{\the\year{}}

\begin{document}
\label{firstpage}
\pagerange{\pageref{firstpage}--\pageref{lastpage}}
\maketitle

\begin{abstract}
Magnetic fields can launch outflows through the Blandford-Znajek (BZ) or the Blandford-Payne (BP) mechanism in accreting black hole systems. Despite the dominant launching mechanism still being unclear, the BP process has been insufficiently studied in general-relativistic magnetohydrodynamic (GRMHD) simulations due to difficulties in generating and maintaining the required large-scale magnetic field geometry. Here, we set up six resistive GRMHD simulations of thin accretion disks with \texttt{BHAC}. The accretion disks in two simulations become magnetically arrested, while the others stay in the standard and normal evolution (SANE) regime. All simulations form jets and winds. The jets are parabolic and collimated by the outer sub-relativistic disk wind; a power law relating the jet width $x$ to the height $z$ as $x\propto z^\mathcal{K}$ yields exponents between 0.52 and 0.64. Among the SANE simulations, the setup with spin parameter $a_* = -0.9375$ has the most efficient jet. For a highly spinning black hole with $a_*=0.9375$, the BZ jet power is $\gtrsim3$ times greater than the disk wind power. The simulated wind regions consist of a highly magnetized region, where the magnetization parameter $\sigma\gg1$, and a less magnetized region. A fast outflow with the potential to reach Lorentz factors up to $\Gamma_\infty\approx25$ propagates in the former region, but because the Lorentz force acts against the disk, it cannot remove a measurable amount of mass from the disk. In the less-magnetized region, a centrifugally driven slower wind efficiently carries away disk material. \end{abstract}




\section{Introduction}
Magnetic fields are a vital component of accretion processes around black holes. In accretion disks, magnetic fields, either dragged inwards or generated through dynamo processes \citep[e.g.,][]{Lubow1994MNRAS,Brandenburg1995ApJ,Spruit2005ApJ,JacqueminIde2024MNRAS}, facilitate the angular momentum transfer through the magnetorotational instability \citep[MRI,][]{Balbus1991ApJ}.

Moreover, magnetic fields can launch outflows, which not only provide another avenue for removing angular momentum but also for a variety of observational phenomena. Magnetic fields can launch outflows through the Blandford-Znajek \citep[BZ,][]{Blandford1977MNRAS} or the Blandford-Payne \citep[BP,][]{Blandford1982MNRAS} mechanism. In the BZ mechanism, magnetic fields threading a spinning black hole extract its rotational energy to power outflows, while in the BP process, magnetic fields threading the accretion disk extract the disk's rotational energy to launch outflows loaded with disk material. These outflows are observed in the form of jets and winds across multiple energy bands \citep[][for a review, see e.g., \citealt{Blandford2019ARA&A..57..467B,MunosDarias2026SSRv..222...39M,Ribeiro2026Univ}]{Russell2019ApJ,Giustini2023A&A...679A..73G}.

There is clear observational evidence for the simultaneous presence of radio jets and mass-loaded winds in radio-loud active galactic nuclei (AGNs), in radio-loud broad-absorption-line quasars, and in black hole X-ray binaries \citep[e.g.,][]{GreggBeckerEtAl2000,MillerBrandtEtAl2006,TombesiSambrunaEtAl2010,TombesiTazakiEtAl2014,Munoz2017MNRAS.465L.124M,SanchezSierras2023A&A...680L..16S}. However, 
which mechanisms are responsible for jet- and wind- launching is still an open question. In particular, assuming BZ and BP processes are responsible for jet- and wind- launching, which process dominates energetically and how the two processes interact remains poorly constrained by observations \citep[e.g.,][]{Russell2013MNRAS.431..405R}. Both launching mechanisms operate close to the black hole, and observationally resolving the actual launch site is feasible only for a few select cases, such as the low-luminosity AGN M87* \citep[which shows no indication for a classical thin-disk or disk-launched outflow however][]{CollaborationAkiyamaEtAl2019,EHTMWLScienceWorkingGroupAlgabaEtAl2021}. 

On the theoretical side, magnetohydrodynamic (MHD) simulations have become a prominent tool for elucidating jet-launching mechanisms. 
However, while properties of BZ-driven outflows, such as their power and dependence on parameters like black hole spin, have been repeatedly studied in general-relativistic MHD (GRMHD) simulations \citep[e.g.,][]{Tchekhovskoy2011MNRAS,Tchekhovskoy2012MNRAS.423L..55T,Morales2018MNRAS,Dhruv2025ApJS}, studies of the BP mechanism have been limited to Newtonian or special relativistic MHD simulations \citep[e.g.,][]{Kato2002ApJ...565.1035K,Zanni2007A&A,Porth2010ApJ...709.1100P}. In GRMHD simulations, generating and maintaining the magnetic field geometry required for BP outflows has proven to be challenging, such that disk winds are often pressure driven in those simulations \citep[][]{HawleyKrolik2006,Vourellis2021ApJ}. 

One way to maintain the magnetic field geometry is a departure from ideal MHD. In ideal MHD, magnetic field lines are frozen into the underlying plasma. Consequently, magnetic flux is dragged inward until it saturates, choking off the accretion process and forming a magnetically arrested disk \citep[MAD, ][]{Narayan2003PASJ}. Resistive MHD lifts the frozen-in condition, which enables the accreting matter to cross magnetic flux surfaces and form a steady-state accretion stream in which inward advection of field lines is balanced by outward diffusion. In simulations, resistivity is either prescribed \citep[e.g.,][]{Striani2016MNRAS,Ripperda2020ApJ,Moran2025ApJ}, or numerical \citep[in ideal GRMHD simulations, e.g.,][]{Ripperda2022ApJ...924L..32R,Jiang2023MNRAS}, the latter resulting from the finite resolution when evolving the MHD equations \citep{Rembiasz2017ApJS..230...18R}. Resistive MHD also allows magnetic field lines to reconnect, which has been connected to flux eruptions observed for Sgr A$^*$ \citep[e.g.,][]{Dexter2020MNRAS,Porth2021MNRAS,Nathanail2022MNRAS,Scepi2022MNRAS}. Hence, non-ideal MHD facilitated by, e.g., resistivity, majorly impacts observations.

Despite the importance of resistivity in accretion, few global resistive GRMHD simulations exist, especially those featuring thin accretion disks capable of launching BP-driven outflows. 
\cite{Vourellis2019ApJ} and \cite{Vourellis2021ApJ} ran resistive GRMHD simulations of thin accretion disks using \texttt{rHARM3D}, which yielded outflows in the form of jets and winds. However, turbulence in their simulated accretion flows prevented BP-driven outflows from forming. Thus, uncertainty remains about the properties of BP winds and their relationship to BZ-driven jets in GRMHD. We aim to address this knowledge gap by running resistive GRMHD simulations of thin accretion disks with BP and BZ-driven outflows. We structure our paper as follows: We outline the simulation setups in section \ref{sec:methods} and detail the results in section \ref{sec:results}. We discuss the results further in section \ref{sec:discussion}, and conclude with section \ref{sec:conclusions}.  

\section{Methods}\label{sec:methods}

\subsection{Numerical setup}\label{subsec:numset}

We run resistive GRMHD simulations with the code \texttt{BHAC} \citep{Porth2017ComAC,Ripperda2019ApJS}. The code solves the conservation equations of energy and momentum and the Maxwell equations via a 3+1 decomposition \citep[e.g.,][]{Rezzolla2013rehy}. For our simulations, we choose modified Kerr-Schild coordinates with $h=0$ and $R_0=0$ \citep{Porth2017ComAC}, which relate the internal coordinates $\anglB{s,\vartheta}$ to $\anglB{r_{KS},\theta_{KS}}$ as 
\begin{align}
    r_{KS}(s) &= e^s\label{eq:intCorS}\\
    \theta_{KS}(\vartheta) &= \vartheta.
\end{align}
Note that in this paper, we will refer to the cylindrical radius as $\rcyl$, and to the spherical radius as $\rsp$. The setup uses a (+,-,-,-) spacetime metric signature and geometric units, where the gravitational constant $G$ and the speed of light $c$ are set to one. We will report our results in code units. Our length, density, and mass units are $r_g=GM/c^2$, $1$, and $c$, respectively, where $M$ is the black hole mass. The time unit is consequently $r_g/c$.

We set up our axisymmetric simulations in two dimensions. Our simulation space covers a radial range of $\squB{1.18,\simeq 10, 000} r_g$ and an angular range of $\squB{0,\pi}$. We use up to three levels of adaptive mesh refinement (AMR), with a base resolution of 512 cells in the radial direction and 192 cells in the radial one. We enforce coarsening of the grid near the polar axes and enforce refinement to the highest level in an angular range of $\approx0.47\pi$ to $\approx0.53\pi$ to ensure that the disk is resolved with an effective resolution of $2048\times768$ cells. The radial boundaries are continuous, and we apply polar symmetry boundaries at the axes. 

We evolve the simulations with an implicit–explicit Runge–Kutta scheme \citep{Palenzuela2009MNRAS.394.1727P,Ripperda2019ApJS}. Primitive variables are recovered through an entropy inversion if needed or if $\beta\leq 10^{-4}$ \citep{Porth2017ComAC}. Vacuum conditions cause numerical problems in GRMHD simulations. Thus, we adjust the densities and pressures during runtime if they drop below density or pressure floors. Densities are rescaled if they become smaller than $10^{-5}\rsp^{-2}$ or $b^2/\sigma_{max}=b^2/25$. Here, $\sigma$ is the magnetization parameter, $\sigma = b^2/\rho$. Pressure values are readjusted in case they are smaller than $10^{-6}\rsp^{-5/2}$, $\squB{\beta_{f}\,b^2/2} = \squB{\anglB{8\times10^{-3}}b^2/2}$ or $10^{-7}\rho$. In addition, we limit velocities such that their Lorentz factor $\anglB{1-v^2}^{-1/2} \leq20$.


\begin{figure*}
    \centering
    \includegraphics[width=0.99\linewidth]{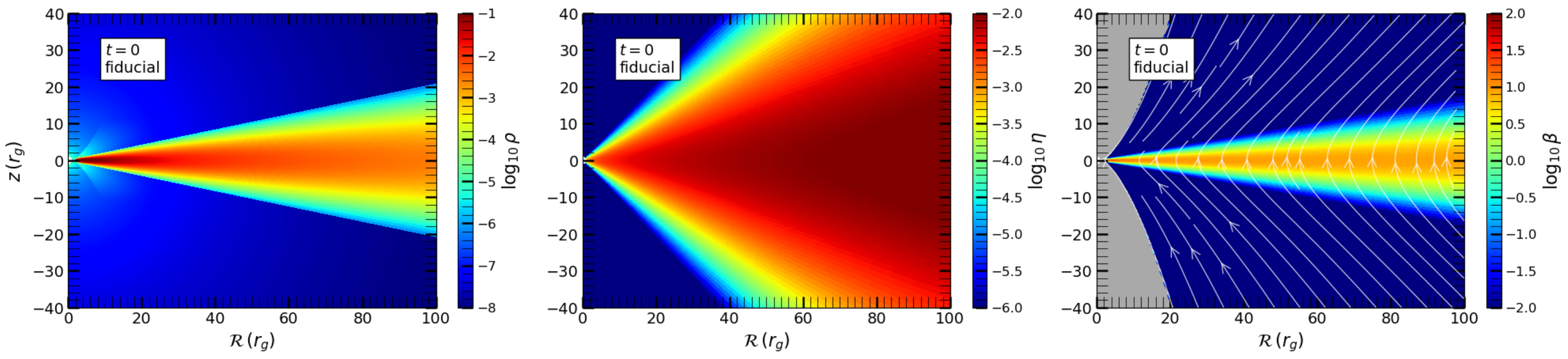}
    \caption{Setup for the fiducial simulation (first row in table \ref{tab:parametersSim}), showing colormaps of the initial density (left panel), resistivity (middle panel), and $\beta = 2p/b^2$ (right panel). The magnetic field lines following the white streamlines in the $\beta$ panel result from the vector potential described by eq. \ref{eq:Aphi}. The region without magnetic field is grayed out.}
    \label{fig:diskSetup}
\end{figure*}
\subsection{Disk initialization}\label{subsec:initdisk}
Assuming the disk is vertically isothermal, the density of the disk has a Gaussian profile (left panel of Fig. \ref{fig:diskSetup}),
\begin{equation}
    \rho(\rsp,\theta) = \rho_0\exp\squB{-\frac{\anglB{\rsp\cos\theta}^2}{2\height^2}},\label{eq:rho}
\end{equation}
where $\rho_0 = \anglB{\rsp\absB{\sin\theta}/r_{ms}}^{-3/2}$ and $\height =\anglB{H/R} \rsp\absB{\sin\theta}$ (for the chosen $H/R$ ratios, see table \ref{tab:parametersSim}).  

The pressure $p$ in the disk depends on the speed of sound $c_s$. We use the thin-disk relation and set $c_s = \height\Omega_K$, with $\Omega_K=1/\squB{\anglB{\rsp\absB{\sin\theta}}^{3/2}+a_*}$ being the Keplerian angular velocity, yielding (eq. \ref{eq:rho})
\begin{equation}
    p(\rsp,\theta) = c_s^2\rho(\rsp,\theta).\label{eq:p}
\end{equation}
The disk velocities are purely Keplerian, i.e., $u^\phi/u^t=\Omega_K$ and the remaining spatial velocity components are set to zero. Choosing the radial exponent $-3/2$ for $\rho_0$ renders the disk near self-similar with constant Mach number at all radii. To distinguish disk material from atmosphere material for later analysis, we additionally evolve one passively advected tracer fluid which is initialized to one inside of the initial disk.

The pressure and density outside of the disk follow $\rho_{f} = 10^{-5}\rsp^{-3/2}$ and $p_{f} = 10^{-6}\rsp^{-5/2}$, respectively. We define a region to be outside of the disk if $\rho \leq\rho_f$ (eq.\ref{eq:rho}), $p\leq p_f$ (eq.\ref{eq:p}), or $\rsp\leq r_{ms}$. Between $r_{ms}\leq \rsp\leq2r_{ms}$, we linearly interpolate the densities and pressure between eq.\ref{eq:rho} and eq.\ref{eq:p} and the density and pressure floor profiles respectively. The marginally stable radius depends on the black hole spin $a_*$ and is 

\begin{align}
    r_{ms} &= M\squB{3+Z_2\mp \sqrt{\anglB{3-Z_1}\anglB{3+Z_1+2Z_2}}}\label{eq:rms}\\
    Z_1  &= 1 +  \anglB{1-a_*^2}^{1/3}\squB{\anglB{1+a_*}^{1/3
    }+\anglB{1-a_*}^{1/3
    }}\\
    Z_2  &= \sqrt{3a_*^2 + Z_1^2}.
\end{align}
The $-$ sign in eq.\ref{eq:rms} applies to positive spins (prograde) and the $+$ sign to negative ones (retrograde). Outside of the disk region, all velocity components are initialized to zero.

We choose a resistivity profile similar to e.g., \cite{Casse2002ApJ} and \cite{Zanni2007A&A},
\begin{equation}
    \eta(\rsp,\theta) = 0.5 \height c_s\exp\squB{-2\anglB{\frac{\rsp\cos\theta}{10\height}}^2},\label{eq:eta}
\end{equation}
which is displayed in the middle panel of Fig.\ref{fig:diskSetup} for the fiducial simulation. The resistivity is floored at $\eta=10^{-6}$ where eq.\ref{eq:eta} yields $\eta<10^{-6}$.  
The broader resistivity profile (eq.\ref{eq:eta}) compared to the initial density profile (compare the left and the middle panel of Fig. \ref{fig:diskSetup}) ensures the disk evolution is governed by anomalous resistivity rather than directly resolved MRI turbulence.

In the absence of radiative cooling, dissipated energy would inevitably cause the disk to heat up and expand vertically. To keep the scale-height under control throughout the time evolution, we hence apply the cooling description introduced by \cite{Noble2009ApJ} \citep[for similar applications, see e.g.,][]{Dhang2025ApJ}. We aim for the temperature (for the chosen $H/R$ ratios, see table \ref{tab:parametersSim})
\begin{equation}
    T_* = \squB{\rsp\sin\theta\anglB{\frac{H}{R}}\Omega_K}^2.\label{eq:temptarget}
\end{equation}
With 
\begin{equation}
    Y \equiv \frac{p}{\rho T_*},
\end{equation}
our cooling function is
\begin{equation}
    L = \xi \Omega_K \frac{p}{\hat{\gamma}-1}\anglB{Y - 1+\absB{Y-1}}\label{eq:lcool},
\end{equation}
where the adiabatic index $\hat{\gamma}$ is $4/3$, and the parameter $\xi$ signifies the ratio of dynamical to thermal timescale; it is set to $\xi=0.1$, which cools sufficiently fast to govern the scale height while allowing the disk to maintain vertical hydrostatic equilibrium.  

\subsection{Magnetic field initialization}
We initialize the magnetic field with a vector potential $A_\flux$ \citep[e.g.,][]{Zanni2007A&A,Vourellis2019ApJ}
\begin{equation}
    A_\flux(\rsp,\theta) =  \left\{\begin{matrix}10^{-6}\frac{\anglB{\rsp\sin\theta}^{3/4} m_p^{5/4}}{\anglB{m_p^2 + \tan^{-2}\theta}^{5/8}}  & A_\flux(\rsp,\phi) > A_\flux\anglB{r_{ms},\pi/2} \\
A_\flux\anglB{\rsp = r_{ms},\theta = \pi/2} &  A_\flux(\rsp,\phi) \leq A_\flux\anglB{r_{ms},\pi/2}\\ \end{matrix}\right.\label{eq:Aphi}
\end{equation}
The poloidial field lines are steeper for smaller constants $m_p$ \citep[compare Fig. 2c and 2d in][]{Dihingia2021MNRAS}. We set $m_p=0.5$ for all simulations. 

The strength of the magnetic field, denoted as $b$ if measured in the fluid frame and $B$ if measured in the Eulerian frame, is parametrized via the plasma-beta $\beta = 2p/b^2$, which we normalize to a target value ranging from 1 to 10 (see table \ref{tab:parametersSim}) at the position $\anglB{\rsp = 10r_{ms},\theta = \pi/2}$. With this normalization procedure, $\beta$ is close to our target $\beta$ along the entire disk midplane (right panel of Fig. \ref{fig:diskSetup}). The gray areas in the rightmost panel of Fig.\ref{fig:diskSetup} are due to the absence of an initial magnetic field (because $A_\flux$ is constant, eq. \ref{eq:Aphi}).

\subsection{Simulation overview}\label{subsec:runSim}
We run seven simulations with parameters and runtimes given in table \ref{tab:parametersSim}. We will refer to one simulation as the fiducial simulation. The other simulations differ by one parameter setting (in bold in table \ref{tab:parametersSim}) to the fiducial one and will be compared against it. All simulation signifiers used to refer to the simulations in this paper are listed in the first column of table \ref{tab:parametersSim}.
\begin{table*}
\begin{tabular}{|c|c|c|c|c|c|c|}
\hline
Simulation signifier  & $\beta$ & $a_*$ & $H/R$ & amount of basic level grid cells [$r\times\theta$]& AMR levels&runtime ($r_g/c$)\\ \hhline{|=======|}
fiducial  & 10 & 0.9375 & 0.05 & [$512\times192$]&3 & 30000 \\ \hline
$\beta = 1$  & \textbf{1} & 0.9375 & 0.05 & [$512\times192$]&3& \textbf{20000} \\ \hline
$\beta = 5$ & \textbf{5} & 0.9375 & 0.05 & [$512\times192$]&3& \textbf{25000} \\ \hline
$a_* = 0.5$  & 10 & \textbf{0.5} & 0.05 & [$512\times192$]&3& 30000 \\ \hline
$a_* = -0.9375$  & 10 & \textbf{-0.9375} & 0.05 & [$512\times192$]&3& \textbf{40000} \\ \hline
$H/R = 0.5$  & 10 & 0.9375 & \textbf{0.1} & [$512\times192$]&3& 30000 \\ \hline
high  & 10 & 0.9375 & 0.05 & \textbf{[}$\mathbf{1024\times384}$\textbf{]} &3& \textbf{20000}\\ \hline
\end{tabular}
    \caption{Summary of simulation parameters. The leftmost column lists the names used for the different simulations. Parameter choices that differ from the fiducial simulation choices are in bold.}
    \label{tab:parametersSim}
\end{table*}

To compare simulations with different spin parameters, the spin dependence of the initial conditions needs to be considered. We chose to normalize the peak value of the initial density profile (eq.\ref{eq:rho}) to one, regardless of $r_{ms}$ (eq.\ref{eq:rms}). However, since the peak is located at $\simeq 2 r_{ms}$ which ranges from $\sim 4 r_g$ (fiducial case) to $\sim 20 r_g$ (counter-rotating case), this  yields larger overall densities for the $a_* = 0.5$ and $a_*=-0.9375$ simulations. Similarly, eq.\ref{eq:Aphi} results in a larger $A_\flux$ as a function of spin at fixed radius. Since this density difference persists until late time, we therefore additionally compute some of the quantifications for the $a_* = 0.5$ and $a_*=-0.9375$ simulations in section \ref{sec:results} with rescaled data.  These datapoints are labeled as 'rescaled' in the figures. In general, this rescaling is possible since our (non-radiative) test-fluid MHD simulations are scale-free in the density.  
Concretely, we divide quantities dependent on $\rho$ or $p$, by 
\begin{equation}
\rescale\equiv\squB{r_{ms,a_{*l}}/r_{ms;0.9375}}^{3/2}\label{eq:rescale},
\end{equation} and magnetic field dependent quantities like $\flux$ by $\sqrt{\rescale}$. 
Here, $a_{*l}$ is either 0.5 or $-0.9375$. The scale factor $\rescale$ is $\approx2.98$ and $\approx8.97$ for $a_*=0.5$ and $a_*=-0.9375$, respectively.   
We provide rescaled density slices along the midplane as an example in the appendix \ref{ap:densInit} which demonstrates that rescaling aligns the densities at large radii between all spin parameters. Dimensionless ratios, such as $\mad$, are not affected by the rescaling, and we expect them to be the least impacted by the spin dependence of the initial conditions. 
We caution that some spin dependence of the setups will persist despite rescaling though. For example, in the parametrization adopted here, the resistivity profile follows from the sound-speed which is sensitive to the spin via the Keplerian angular velocity $c_s=\mathcal{H}\Omega_K(a_*)$.  

\subsection{Analysis} \label{subsec:analysis}
At a given radius, we measure the mass flow rate $\mdot$ and the magnetic flux $\flux$ as \citep{Porth2017ComAC},
\begin{align}
    \mdot =& -\int_0^\pi \rho u^r \sqrt{-g}\, 2\pi\,d\theta \label{eq:mdot}\\
    \flux =& \frac{1}{2}\int_0^\pi \absB{B^r} \sqrt{-g}\, 2\pi\,d\theta. \label{eq:phi}
\end{align}
Equations \ref{eq:mdot} and \ref{eq:phi} evaluated at the black hole horizon can be combined to the MADness parameter,
\begin{equation}
    \mad = \left.\frac{\flux}{\sqrt{\mdot}}\right\rvert_{r=r_H},\label{eq:mad}
\end{equation}
which measures how magnetically arrested a disk is. In ideal GRMHD simulations, a thick disk becomes MAD if $\mad\approx15$ \citep[e.g.,][]{Tchekhovskoy2011MNRAS,Liska2020MNRAS}. Simulated thin accretion disks tend towards lower MADness parameters \citep[e.g.,][]{Avara2016MNRAS,Liska2022ApJ,Scepi2024MNRAS,Lowell2025PhRvD}.

The mass flow rate can be divided into the inflowing accretion rate and the mass outflow rate \citep[e.g.,][]{Fragile2026MNRAS}
\begin{align}
    \mdotin = &-\int_0^\pi \min\anglB{u^r,0} \rho\sqrt{-g}\, 2\pi\,d\theta \label{eq:mdotin}\\
    \mdotout = &\int_0^\pi \max\anglB{u^r,0} \rho\sqrt{-g}\, 2\pi\,d\theta \label{eq:mdotout}.  
\end{align}
Part of the mass outflow rate is unbound, denoted as $\mdotun$, if the Bernoulli parameter \citep[e.g.,][]{Zhang2025ApJ}
\begin{equation}
    Be = -\anglB{\rho+\frac{\hat{\gamma} p}{\hat{\gamma}-1} + b^2}\frac{u_t}{\rho}-1\label{eq:bern}
\end{equation}
is greater than zero. 

Unbound matter with $Be>0$ (eq. \ref{eq:bern}) belongs either to the jet or the disk wind, while bound material counts towards the disk. The $Be=0$ contour is visible in Fig. \ref{fig:theEndRho} for the fiducial simulation at the end of its runtime and follows the magenta dashed line (we will discuss Fig. \ref{fig:theEndRho} in more detail in section \ref{subsec:simev}). To separate the jet from the wind region, we calculate a time-dependent vector potential $\Tilde{A}_\flux$
\begin{equation}
    \Tilde{A}_\flux (r,\theta,t) = \int_0^\theta B^r \sqrt{-g}\, 2\pi\,d\theta,\label{eq:aphiTime}
\end{equation}
where we write $\Tilde{A}_\flux$ rather than $A_\flux$ to differentiate it from the initial vector potential profile described by eq. \ref{eq:Aphi}. We define the jet region to be bounded by the last magnetic field line emerging from the black hole, which follows a $\Tilde{A}_\flux =\flux|_{r=r_H}$ contour (dotted magenta lines in Fig. \ref{fig:theEndRho}). Areas beyond this contour but with $Be > 0$ are part of the wind. As a consequence of this classification, we will refer to outflows powered by the BZ process as jets and BP-driven outflows as winds in this paper. However, as we will show in section \ref{subsec:windV} and discuss further in section \ref{sec:whyWind}, part of the BP-driven outflow strongly resembles a jet and could be attributed to it in observations.

To analyze the wind and jet powers, we decompose the stress-energy tensor into electromagnetic (EM), free particle (PAKE), and enthalpy (EN) terms. In this work, we report the $T^r_t$ components as a function of radius \citep{McKinney2012MNRAS,Qian2017ApJ},
\begin{equation}
    \begin{aligned}
    {T^{EM}}^{r}_t = -\int&\sqrt{-g}  \left[\anglB{b^2 + e^2}u^r u_t-b^r b_t - e^r e_t \right.\\ &\left. - u_\lambda e_\beta b_\kappa \anglB{u^r\anglB{-g}^{-1/2}\eta_t^{\lambda\beta\kappa} + u_t \anglB{-g}^{-1/2}\eta^{r\lambda\beta\kappa}}\right]\,2\pi\,d\theta,\label{eq:Tem}
\end{aligned}
\end{equation}
\begin{equation}
    {T^{PAKE}}^{r}_t = -\int\sqrt{-g}\squB{\anglB{u_t+1}\rho u^r}\,2\pi\,d\theta,\label{eq:Tpake}
\end{equation}
\begin{equation}
    {T^{EN}}^{r}_t = -\int\sqrt{-g}  \squB{\frac{\hat{\gamma} p}{\hat{\gamma}-1}u^r u_t}\,2\pi\,d\theta.\label{eq:Ten}
\end{equation}
In eq. \ref{eq:Tem}, $\eta$ is spatial Levi-Civita symbol \citep{Ripperda2019ApJS}, and $e$ is the electric field in the fluid frame. The sum of eq.\ref{eq:Tem}-\ref{eq:Ten} is the total energy flux.



The total energy flux, i.e., the power, will be the sum of the stress-energy tensor components (eq.\ref{eq:Tem}-\ref{eq:Ten}). However, when measuring the total energy flux of the jet at the black hole horizon, we will only use the EM component \citep[as done in e.g.,][]{Selvi2024ApJ...968L..10S}. 
 
In this work, we will refer to the total energy flux of the jet as $P^{BZ}$, and the total energy flux of the wind as $P^{BP}$. To present the wind and jet energetics, we sum the northern and southern wind/jet together.

Because of the energy flux, matter can be transported towards large distances. We calculate the Lorenz factors reached at infinity as \citep[compare with e.g.,][]{McKinney2006MNRAS.368.1561M,Tchekhovskoy2010NewA...15..749T}
\begin{equation}
    \Gamma_\infty = -\left.\frac{\anglB{{T^{EM}}^{r}_t + {T^{EN}}^{r}_t + {T^{PAKE}}^{r}_t - \dot{M}}}{\dot{M}}\right|_{\mathrm{not\,integrated}},\label{eq:gammaInf}
\end{equation}
where we use the non-integrated expressions of eq. \ref{eq:mdot} and \ref{eq:Tem}-\ref{eq:Ten} (i.e. without $\int \,d\theta$) to retain the angular distributions of the Lorentz factors.

When plotting time-averaged quantities, we use error bars. If the quantity is not a fraction, the error bars will be one standard deviation. Unless specified otherwise, the error bars of fractions $x/y$, such as $\mad$ (eq. \ref{eq:mad}), are calculated as 
\begin{equation} \mathrm{error}_{x/y}=\Bar{x}/\Bar{y}\sqrt{\anglB{\sigma_x/\Bar{x}}^2+\anglB{\sigma_y/\Bar{y}}^2},\label{eq:propError}
\end{equation}
where $\Bar{x}$ and $\Bar{y}$ are the averages and $\sigma_x$ and $\sigma_y$ the corresponding standard deviations.

To compute time-averages, we always use the last 5000 $r_g/c$ of a respective simulation (runtimes are listed in table \ref{tab:parametersSim}).


\section{Results}\label{sec:results}
Section \ref{sec:results} outlines our results. We will first describe the evolution of the simulations, with a focus on the mass flow rate and magnetic flux accumulated at the black hole horizon (section \ref{subsec:simev}). Section \ref{subsec:massflow} examines the mass flow as a function of radius. In section \ref{subsec:jetprop}, we analyze the jet properties. Section \ref{subsec:windprop} details the properties of the wind region, which we define as the region between the outermost magnetic field line still anchored at the black hole and the $Be=0$ contour. As a consequence, we refer to BZ-driven outflow as a jet and to BP-driven outflow as a wind (although part of the BP-driven outflow may also be regarded as a jet, see section \ref{subsub:obshighwind}).
\subsection{Simulation evolution}\label{subsec:simev}
\begin{figure*}
    \centering
\includegraphics[width=0.9\linewidth]{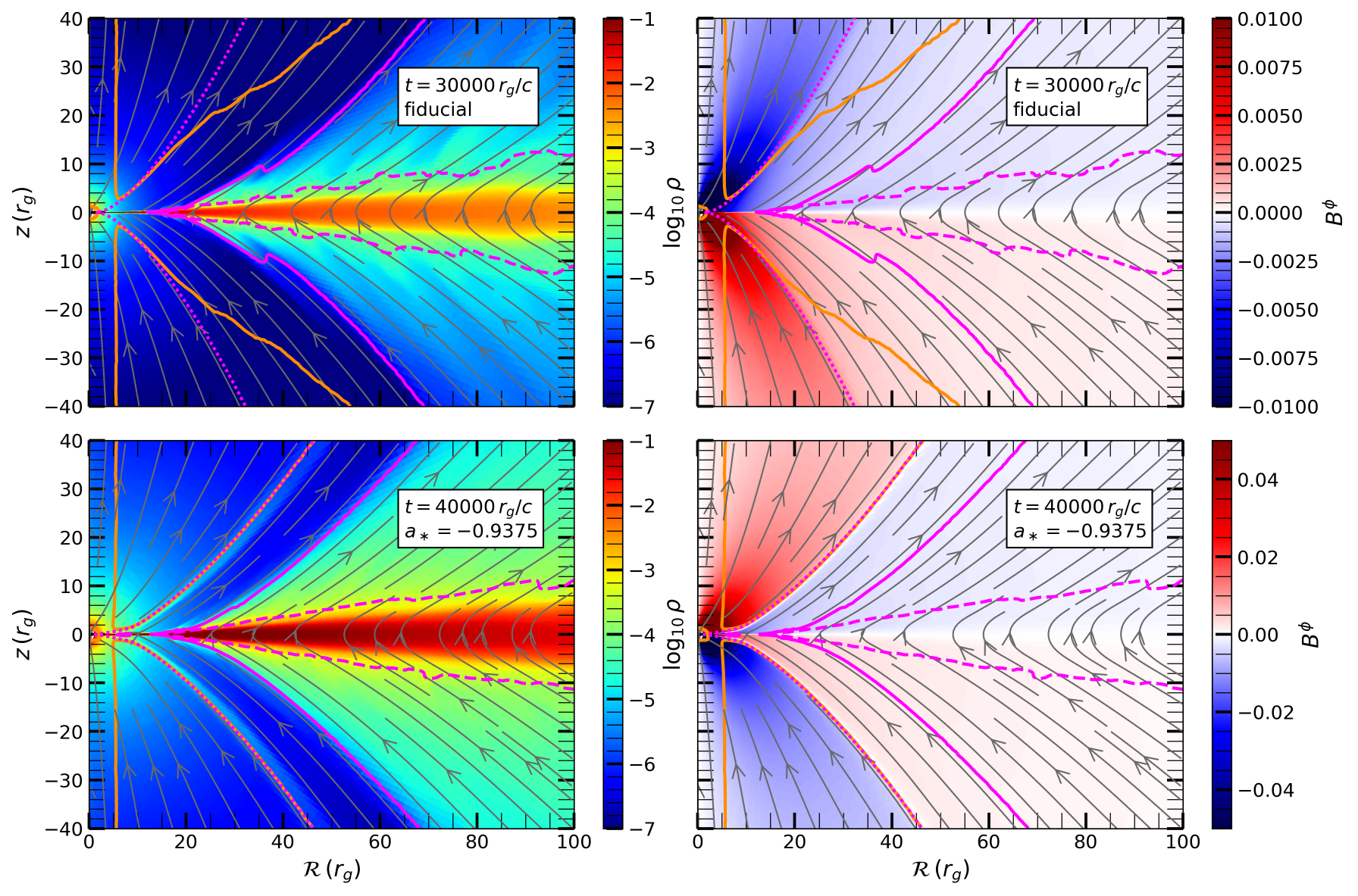}
    \caption{Left column: Density map of the fiducial simulation at $t=30000r_g/c$ (top panel) and of the counterrotating simulation at $t=40000r_g/c$ (bottom panel). The $Be=0$ contour line (magenta dashed line, eq.\ref{eq:bern}) confines the bound material of the disk. A disk wind removes unbound disk material. The region close to the black hole is highly magnetized with $\sigma>1$ (the magenta solid line follows the $\sigma=1$ contour). However, part of this $\sigma>1$ region still counts towards the wind, as the magenta dotted line traces the last magnetic field line emerging from the black hole, which bounds the jet region. The gray arrows follow the magnetic field lines, and the orange lines represent the light cylinder boundary. Right column: $B^\phi$ at the final timestep of the fiducial (top panel) and the counterrotating simulation (bottom panel). The lines are the same as for the left column. The opposite polarities between the jet and wind regions in the $a_*=-0.9375$ simulation create a current sheet through which matter flows outwards.}
    \label{fig:theEndRho}
\end{figure*}
Soon after the simulation starts, a disk wind emerges close to the black hole\footnote{A video of the fiducial simulation is available here \url{https://www.youtube.com/shorts/DK0sWnmahMg}.}. An increasing magnetic pressure and the cooling function (eq.\ref{eq:lcool}) concentrate the disk densities towards the midplane. Although there is no initial magnetic field at the poles and the initial magnetic field has no $B_\phi$ component (rightmost panel of Fig. \ref{fig:diskSetup}, eq. \ref{eq:Aphi}), the magnetized region and $B_\phi$ grow quickly. The left column of Fig. \ref{fig:theEndRho} shows a density map, along with magnetic field lines (gray streamlines) for the fiducial simulation (top panel) and the $a_*=-0.9375$ simulation (bottom panel) at their respective last time steps. Except for the $a_*=-0.9375$ simulation, $B_\phi < 0$ in the entire upper (northern) hemisphere and positive in the entire lower (southern) hemisphere (right column). For the retrograde simulation, this partition is reversed in the jet region, which is bounded by the last emerging field line anchored at the black hole (magenta dotted lines). The light cylinder (orange line, section \ref{ap:lightCyl}) is shown for reference as it marks the region where electric fields play a role in the trans-field force balance. Both jet and disk winds cross the light cylinder in the fiducial simulation \citep[compare with e.g., ][]{Porth2010ApJ...709.1100P}.
 Most material is confined within the accretion disk (within the $Be=0$ magenta dashed contour line, eq.\ref{eq:bern}). The large-scale structured magnetic field drives a disk wind and removes matter from the disk. The inner-wind region is highly magnetized, extending beyond the solid magenta $\sigma = 1$ contour.

\subsubsection{Mass flow rate and magnetic flux evolution}
\begin{figure*}
    \centering
    \includegraphics[width=0.99\linewidth]{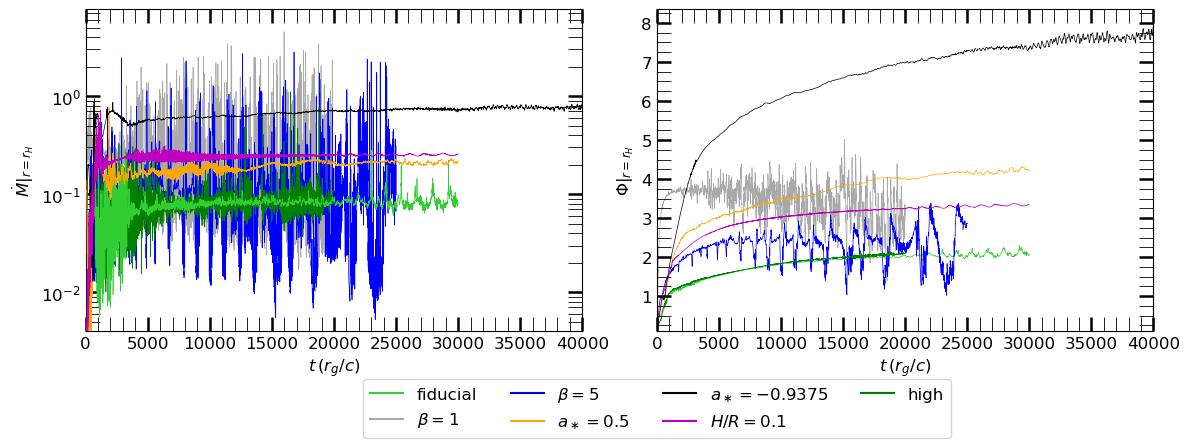}
    \caption{Left panel: Mass flow rate (eq. \ref{eq:mdot}) evaluated at the horizon as a function of time. Right panel: Same as the left panel, but for the magnetic flux (eq. \ref{eq:phi}). As the simulations start, the \mdotname increase quickly, while the magnetic fluxes grow over longer time scales. The $\mdot$ and $\flux$ evolutions of the $\beta=1$ (gray line) and $\beta=5$ (blue line) simulations exhibit strong fluctuations due to magnetic reconnection.}
    \label{fig:MdotPhi}
\end{figure*}

As the simulations begin, accretion sets in and drags magnetic flux inward. Fig. \ref{fig:MdotPhi} shows the time evolution of the \mdotname (eq. \ref{eq:mdot}, left panel) and the magnetic flux (eq. \ref{eq:phi}, right panel) evaluated at the black hole horizon. The $\mdot$ and $\flux$ evolution of the fiducial (green lines) and the double-resolution simulation (dark green lines) closely resemble each other, showing that our chosen standard resolution is sufficient. The \mdotname increases quickly and reaches $95\%$ of its late time value within $t\lesssim 800 r_g/c$ for all simulations. However, the time-averaged $\dot{M}$ can deviate significantly from the instantaneous value, especially in the $\beta=1$ (gray line) and $\beta=5$ (blue line) simulations. These two simulations exhibit frequent magnetic flux eruptions and become MAD, while the others remain in the Standard and Normal Evolution (SANE) regime. 

The flux eruptions appear as strong fluctuations in $\flux$ at the horizon for the MAD simulations  (gray and blue lines in the right panel of Fig. \ref{fig:MdotPhi}). In the SANE simulations, the ratios of the standard deviation and average of $\flux$, $\sigma_\flux/\Bar{\flux}$, yield $\lesssim0.03$. In contrast, $\sigma_\flux/\Bar{\flux}$ of the $\beta=1$ and $\beta=5$ simulations are $\approx0.2$ and $\approx0.25$, respectively, as calculated for the last $5000r_g/c$ of the runtimes.

\begin{figure*}
    \centering
    \includegraphics[width=0.95\linewidth]{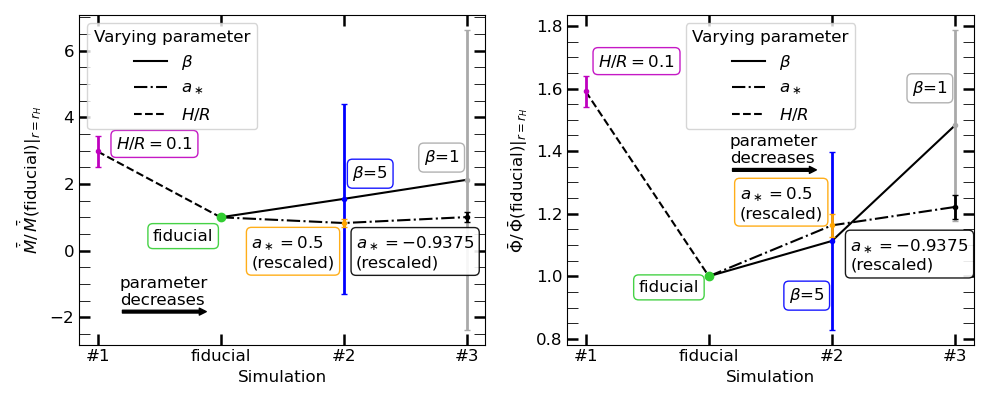}
    \caption{Left panel: Ratios between the time-averaged mass flow rates of a simulation and the fiducial one. The averages are calculated for the last $5000r_g/c$ of a respective simulation and measured at the horizon. Ratios connected by lines belong to simulation setups differing by only one parameter (see numbers next to dots). Moving from left to right, the differing parameter, i.e., $\beta$ (solid line), $a_*$ (dash-dotted line), or $H/R$ (dashed line), decreases. The time-averaged mass flow rates of the $a_*=0.5$ (orange error bar) and $a_*=-0.9375$ (black error bar) simulations are divided by $\lambda$ (eq.\ref{eq:rescale}). Right panel: Same as the left panel, but for the magnetic fluxes. The time-averaged magnetic fluxes of the $a_*=0.5$ and $a_*=-0.9375$ simulations are divided by $\sqrt{\lambda}$.}
    \label{fig:phiTrends}
\end{figure*}

The trends in mass flow rates and fluxes at the horizon, time-averaged over the last $5000 r_g/c$ of a respective simulation, are summarized in Fig. \ref{fig:phiTrends}. The left panel of Fig. \ref{fig:phiTrends} shows the ratios of the time-averaged mass flow rates, normalized to the same quantity of the fiducial simulation. Lines connect data points of setups that differ by only one simulation parameter as indicated. To account for the spin-impact on the initial conditions for the $a_*=0.5$ and $a_*=-0.9375$ simulations, we divide $\Bar{\mdot}$ and $\Bar{\flux}$ by $\rescale$ (eq.\ref{eq:rescale}) and $\sqrt{\rescale}$, respectively.

The time-averaged \mdotname of the fiducial simulation is $\approx0.08$ (in code units). In viscous thin-disk accretion theory, $\mdot\sim H^2$. Thus, doubling the scale height to 0.1 would lead to a four-fold increase in $\dot{M}$. When angular momentum-removal via a disk wind is a major component of the accretion process, the expectation on the scale-height dependence is less clear. However, as the data shows, (purple error bar in left panel), doubling the scale height also increases the \mdotname by $\approx3$ times.

The magnetic field strength affects the \mdotname through its impact on angular momentum transport. Magnetic torques reduce the specific angular momentum of the accreting material, accelerating infall \citep[e.g.,][]{Blandford1982MNRAS,Zanni2007A&A}. Hence, decreasing $\beta$ (solid line) from 10 (fiducial simulation) to $5$ (blue error bar) and $1$ (grey error bar) increases the $\mdot$ ratio to $\approx1.6$ and $\approx2.1$, respectively \footnote{Note that accretion in our simulations does not depend on the strength of the MRI because our large resistivities (eq. \ref{eq:eta}) suppress it in the disk \citep[e.g.,][]{Begelman2023MNRAS}.}. The $\mdot$ enhancements are not higher because flux eruptions cause frequent drops in $\mdot$ (see also the left panel of Fig. \ref{fig:MdotPhi}).

The mass flow rate is almost insensitive to spin (dot-dashed line) after dividing by $\rescale$. Lowering the spin from $a_*=0.9375$ to $a_*=0.5$ (orange error bar) decreases the $\mdot$ ratio slightly to $\approx0.8$ (the unscaled ratio is $\approx2.5$). A counterrotating setup with $a_*=-0.9375$ (black error bar) marginally enhances $\mdot$ compared to the fiducial setup; the ratio is $\approx1.01$ (the unscaled ratio is $\approx9.0$). Our simulations therefore show a different trend of $\mdot$ as a function of spin than thick-disk MAD simulations, where a counterrotating setup led to a more significant $\mdot$ enhancement \citep[e.g.,][]{Tchekhovskoy2012MNRAS.423L..55T,Narayan2022MNRAS}. 

The magnetic flux at the horizon follows the same trends as the mass flow rates (right panel of Fig. \ref{fig:phiTrends}). The two panels of Fig. \ref{fig:phiTrends} are connected since accretion drags $\flux$ inwards. However, the $\mdot$ enhancements (left panel of Fig. \ref{fig:phiTrends}) translate into smaller $\flux$ enhancements. While magnetic-field diffusion contributes to the smaller enhancement, it is also consistent with the scaling of the maximum flux with \mdotname of $\Phi\propto \sqrt{\mdot}$ obtained in the MAD regime. The flux ratios of the $a_*=0.5$ and $a_*=-0.9375$ simulations might be greater than their $\mdot$ ratios due to some remaining spin-dependence from the initial conditions despite dividing by $\sqrt{\rescale}$.  

\subsubsection{MADness parameter}
\begin{figure*}
    \centering
    \includegraphics[width=0.99\linewidth]{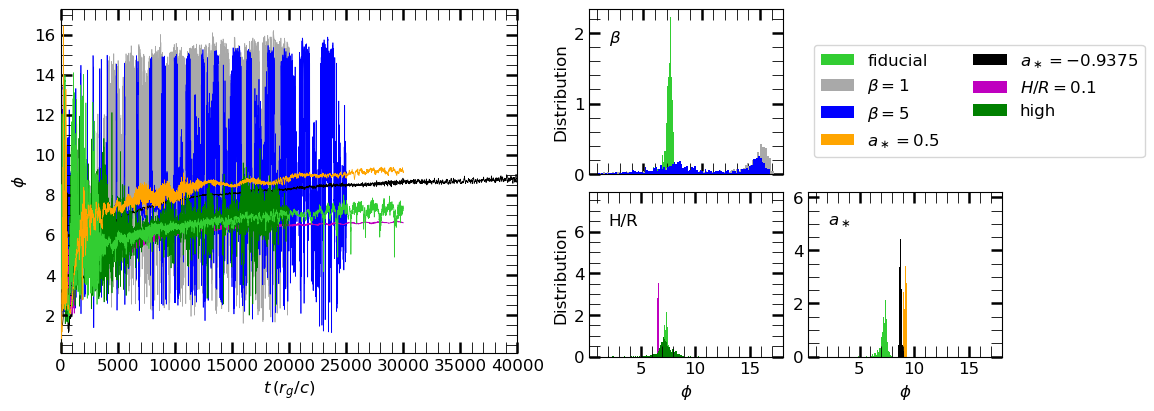}
    \caption{Left large panel: Time evolution of the MADness parameter (eq. \ref{eq:mad}). Right small panels: Distributions of $\mad$ over the respective last $5000r_g/c$. In each subpanel, the simulation setups differ by the parameter written in the respective top-left corner (same groupings as in Fig. \ref{fig:phiTrends}). While $\mad$ varies little in the SANE simulations, the MADness parameter distributions are broader and skewed towards high $\mad$ for the $\beta=1$ and $\beta=5$ simulations (grey and blue distributions in upper small panel).}
    \label{fig:madDist}
\end{figure*}

The \mdotname and magnetic flux measured at the horizon yield the MADness parameter (eq. \ref{eq:mad}) evolution shown in the large left panel of Fig. \ref{fig:madDist}. The MADness parameter visibly remains $\lesssim15$, consistent with other thin-disk simulations \citep[e.g.,][]{Avara2016MNRAS,Liska2019MNRAS,Scepi2024MNRAS,Lowell2025PhRvD}. The right small panels display the $\mad$ distributions for the last 5000 $r_g/c$ of a respective simulation (see table \ref{tab:parametersSim} for the runtimes). Distributions are grouped like in Fig. \ref{fig:phiTrends} to highlight how $\mad$ changes as one simulation parameter (written in top-left corners) varies. The distribution of the high-resolution simulation (dark green) is plotted in the small lower-left panel; its distribution and time evolution closely resemble those of the fiducial simulation (green). While $\mad$ averages to $\approx7.2$ over the last 5000 $r_g/c$ of the fiducial simulation, the MADness parameters of the $\beta = 1$ (gray) and the $\beta=5$ (blue) simulations are distributed over a larger range and show a main peak at $\phi\simeq 15$ (top small panel). Thus, our MAD simulations distinguish themselves from our SANE ones through their broader, negatively skewed $\mad$ distributions.


Consequently, the $\mad$ distributions shown in the other panels are significantly more symmetrical and narrower. The $H/R=0.1$ simulation (purple, small bottom left panel) has with an average $\mad\approx6.6$ and a standard deviation of $\approx3\times10^{-2}$ the narrowest distribution. Thus, our thicker disk setup has a smaller MADness parameter compared to the fiducial case.

In the simulation with the retrograde black hole (black, bottom-right small panel), the average MADness parameter is $\approx8.7$ with a standard deviation of $\approx0.1$. The simulation with $a_*=0.5$ (orange) tends to have higher $\flux/\sqrt{\mdot}$ values, averaging $\approx9.2$ in its last $5000r_g/c$ time interval. Therefore, the $\mad$ averages of our thin SANE simulations follow a different spin relationship than MAD (thick and thin) or SANE thick-disk simulations, where $\mad$ was smallest in the retrograde setups \citep[e.g.,][we are unaware of reported $\mad$ measurements for thin SANE disks]{Tchekhovskoy2012MNRAS.423L..55T,Narayan2022MNRAS,Dhang2025ApJ,Dhruv2025ApJS}.

\subsection{Mass flow}\label{subsec:massflow}
\begin{figure*}
    \centering
    \includegraphics[width=0.9\linewidth]{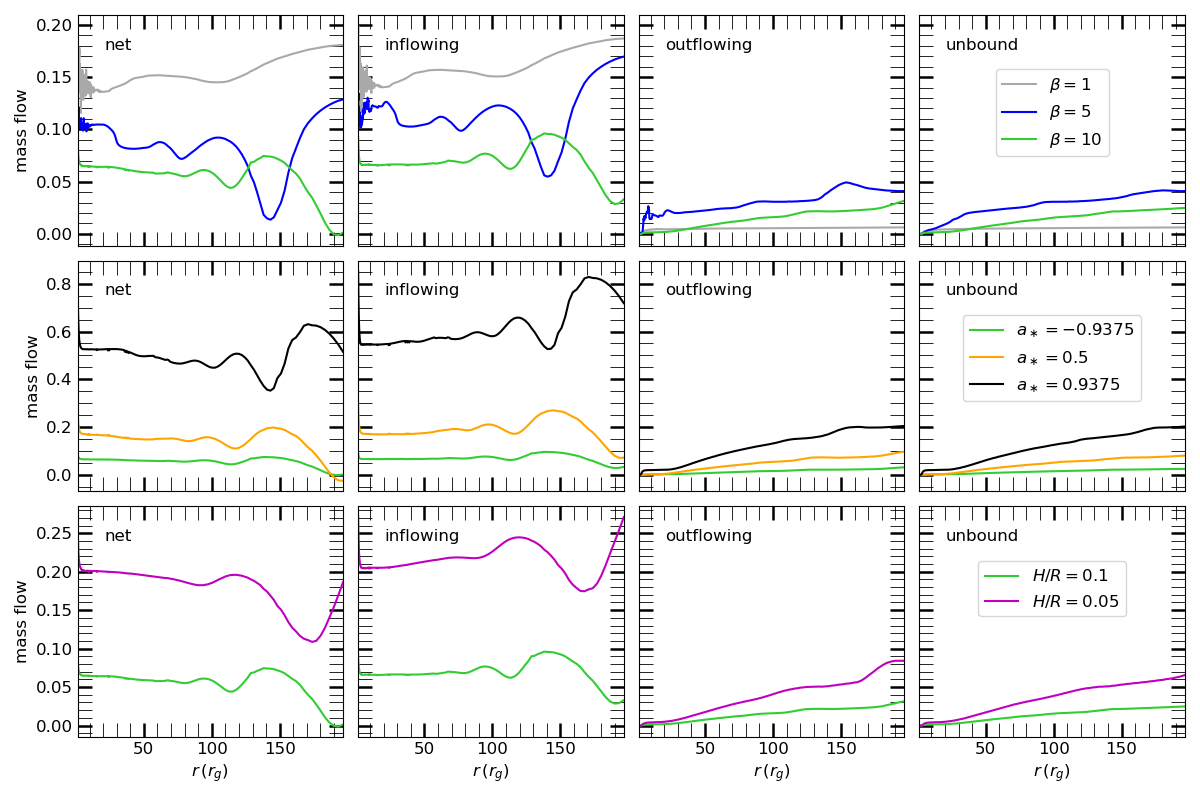}
    \caption{Mass flow rate (eq.\ref{eq:mdot}, first column), subdivided into the inflowing mass accretion rate (second column, eq. \ref{eq:mdotin}), the mass outflow rate $\mdotout$ (third column, eq.\ref{eq:mdotout}), and the unbound mass flow rate (last column). All radial profiles are time averaged over the last $5000r_g/c$ of a respective simulation. Each row shows profiles of the same simulations (as denoted by the legends in the last column).}
    \label{fig:massMovement}
\end{figure*}

Fig. \ref{fig:massMovement} shows the \mdotname as a function of radius, time averaged over the last $5000r_g/c$ of a respective simulation. The first column shows $\mdot$, which can be subdivided into the inflowing mass accretion rate $\mdotin$ (second column, eq. \ref{eq:mdotin}) and the mass outflow rate $\mdotout$ (third column, eq.\ref{eq:mdotout}). A portion of $\mdotout$ is unbound (last column). Note that $\mdotin$ can exceed the \mdotname since eq.\ref{eq:mdot} for $\mdot$ integrates over regions with positive (i.e., outflow) and negative $u^r$ (i.e., inflow). The strong variability of the $\beta=1$ (gray lines, first row) and $\beta=5$ (blue lines, first row) simulations translates into jitter of the time-averaged mass movement profiles at $r\lesssim 10r_g$ in contrast to the smooth ones of the SANE simulations. The transient radial variations of $\mdot$ at larger radii result from accumulations of matter and magnetic flux. These transient disk bulges exhibit enhanced outflows, decreasing the $\mdot$. For instance, the disk of the counterrotating simulation bulges at $r\approx150 r_g$. At this radius, the ratio between $\mdot_{out}$ (eq. \ref{eq:mdotout}) and $\mdot_{in}$ (eq. \ref{eq:mdotin}) is with $\approx0.33$ the greatest for this simulation. However, these matter accumulations are not stationary but slowly drift outwards over time. Thus, $\mdot$ has reached a near-stationary state only within $\sim100r_g$ in our simulations.

While the time-averaged inflowing accretion rate (second column) dominates the \mdotname (first column), the mass outflow rate (third column) increases radially. At $r\approx100r_g$, $\mdot_{out}/\mdot\gtrsim0.21$ for all simulations except for the one with $\beta=1$ (grey lines in top row), which shows almost zero mass ejection from the disk. We will discuss the failure of mass loading by disk winds further in section \ref{subsub:lackmassloading}. Most of the outflowing material is unbound (rightmost column), making up $\gtrsim93\%$ of the outflowing mass at $r\approx100r_g$ for all simulations. Thus, if mass loading by disk winds succeeds, the simulated disk winds are an efficient avenue for removing disk material (see also section \ref{subsec:windV}).

\subsection{BZ outflow}\label{subsec:jetprop}
\subsubsection{Geometric properties}
\begin{figure*}
    \centering
    \includegraphics[width=0.9\linewidth]{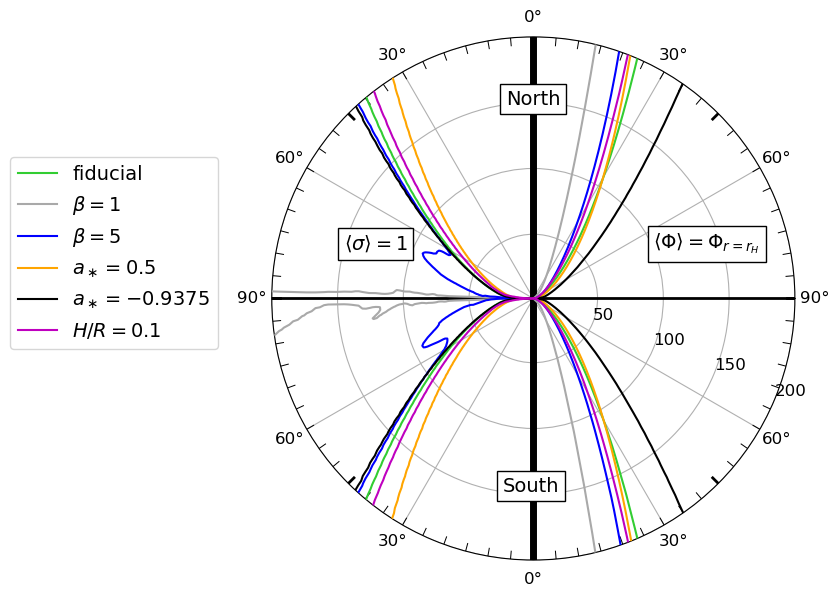}
    \caption{Left side: Contour lines of $\triB{\sigma}=1$, averaged over the last $5000r_g/c$ of a simulation. Right side: same as the left, but showing the outermost magnetic field lines still anchored at the black hole. The field lines follow the contours of $\Bar{\Tilde{A}}_\flux = \flux_{r=r_H}$ (eq. \ref{eq:aphiTime}).}
    \label{fig:sigmaFieldLineContour}
\end{figure*}

A highly magnetized region forms in all simulations (black solid line in Fig. \ref{fig:theEndRho}). The left side of Fig. \ref{fig:sigmaFieldLineContour} shows the contour lines of $\triB{\sigma}=1$, where $\sigma$ has been averaged over the last $5000 r_g/c$. The time-averaged contour lines for the $\beta=1$ (grey lines) and $\beta=5$ (blue lines) simulations show several bends due to their greater time variability. However, as already visible in Fig. \ref{fig:theEndRho}, the $\triB{\sigma}=1$ does not tightly confine the jet (comparing the solid white and magenta lines). The right side of Fig. \ref{fig:sigmaFieldLineContour} shows the time-averaged outermost magnetic lines still anchored at the black hole, which follow the contour of $\Bar{\Tilde{A}}_\flux = \flux_{r=r_H}$ (eq. \ref{eq:aphiTime}). The jets in the $\beta=1$ and $\beta=5$ simulations are the slimmest because they are confined by the strong magnetic flux that has been expelled from the black hole through flux eruptions.

Besides the jets in the $\beta=1$ (grey lines), $\beta=5$ (blue lines) and $a_*=-0.9375$ (black lines) simulations, the time-averaged jet opening angles are similar. At $r\approx50r_g$, the opening angles of the northern jet as measured from the pole differ by less than 7 degrees between the $a_*=0.5$ ($\approx41.5^\circ$, orange lines), the fiducial ($\approx39.1^\circ$, green lines) and the $H/R=0.1$ ($\approx35.8^\circ$, purple lines) lines. Because the jet shapes differ, the trend in jet broadness between the simulations changes at larger radii. At $r\approx200r_g$, the jet of the fiducial simulation is with $\approx23.4^\circ$ broader than the jet of the $H/R=0.1$ ($\approx21.2^\circ$) simulation due to a thinner disk \citep[e.g.,][]{Liska2019MNRAS}. The fiducial simulation is also less collimated than the $a_*=0.5$ one, whose time-averaged opening angle measures $\approx22.0^\circ$ at $r\approx200r_g$. The difference in jet collimation between the fiducial and the $a_*=0.5$ simulation cannot be attributed to a smaller MADness parameter, as in the MAD simulations by \cite{Narayan2022MNRAS}, but determining the exact cause of the difference is beyond the scope of this work. The counterrotating simulation has the broadest jet. At $\approx200 r_g$, the time-averaged opening angle of its northern jet is $\approx35^\circ$, which results from the large amount of flux accumulated at the black hole (see right panels of Fig. \ref{fig:MdotPhi}, \ref{fig:phiTrends}). Besides the different dependence of collimation on spin, it is thus interesting to note that the jets in our simulations are similarly collimated as those in MAD simulations \citep[e.g.,][]{Narayan2022MNRAS,Dhruv2025ApJS}, despite the absence of a thick disk.

The jet shapes responsible for the opening angles can be fitted with a powerlaw relating the jet width $x$ to the height $z$ as $x\propto z^\mathcal{K}$\citep[e.g.,][]{Kovalev2020MNRAS,Narayan2022MNRAS}. Observational studies find AGN jet shapes shift from parabolic, corresponding to $\mathcal{K}=0.5$, to conical with $\mathcal{K}=1$ at large radii $\sim 10^7 r_g$, which are however far beyond our simulation boundary \citep[e.g.,][]{Kovalev2020MNRAS,Lee2025}. When fitting the northern jet boundaries (right side of Fig. \ref{fig:sigmaFieldLineContour}) with $x=\mathcal{A}z^\mathcal{K}$ for $r\leq200r_g$, we obtain exponents ranging from $\mathcal{K}\approx0.52$ for the $a_*=0.5$ simulation to $\mathcal{K}\approx0.64$ for the $\beta=1$ simulation (see table \ref{tab:fitPowerlaw}). Hence, all jets in our simulations have a more parabolic shape which meets observational expectations.  

\begin{table}\centering
\begin{tabular}{|c|c|c|}
\hline
Simulation signifier & $\mathcal{A}\,$&$\mathcal{K}\,$ \\ 
\hhline{|===|}
fiducial& 3.6 & $0.6$ \\ \hline
$\beta = 1$& 1.6&  $0.64$ \\ \hline
$\beta = 5$&3.3 & $0.57$ \\ \hline
$a_* = 0.5$&4.9 & $0.52$ \\ \hline
$a_* = -0.9375$&5.2 & $0.60$ \\ \hline
$H/R = 0.1$&3.2 & $0.60$
\end{tabular}   
\caption{Amplitudes and exponents for fitting the time-averaged jet boundaries for $r\leq200r_g$ (right side of Fig. \ref{fig:sigmaFieldLineContour}). The fits $x=\mathcal{A} z^\mathcal{K}$, where $\mathcal{A}$ is the amplitude, $x$ is the jet width and $z$ is the height, yield exponents of $\mathcal{K}\approx0.6$. With $\mathcal{K}$ being closer to 0.5 than 1, the jets' shapes are more parabolic than cylindrical \citep[e.g.,][]{Pushkarev2017MNRAS.468.4992P}.}
    \label{tab:fitPowerlaw}    
\end{table}

\subsubsection{Jet power and velocities}
The jet shape (right side of Fig. \ref{fig:sigmaFieldLineContour}) is related to the magnetic field geometry, which can be connected to the jet power. Up to sixth order in $\Omega_H = a_*/\anglB{2r_H}$, \cite{Tchekhovskoy2010ApJ} find 
\begin{equation}
    P^{BZ}\approx k\flux_{jet}^2\anglB{\Omega_H^2 + 1.38\Omega_H^4 -9.2\Omega_H^6},\label{eq:jetPowerTch}
\end{equation}
The magnetic flux $\flux_{jet}$ is measured at the black hole horizon in the jet region (see sec.\ref{subsec:analysis} for the definition). The constant $k$ is a function of the magnetic field geometry and is expected to be $\approx\anglB{6\pi}^{-1}$ for a split-monopole configuration and $\approx0.044$ for a paraboloidal one \citep{Tchekhovskoy2010ApJ}.

\begin{figure*}
    \centering
    \includegraphics[width=0.9\linewidth]{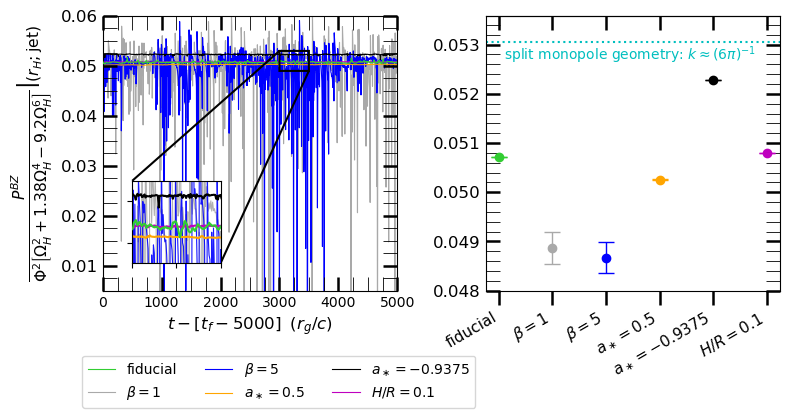}
    \caption{Left panel: Jet power divided by $\flux^2\anglB{\Omega_H^2 + n\Omega_H^4 + m\Omega_H^6}$. All quantities are measured in the total jet region at the black hole horizon. The time is rescaled to show the last 5000 $r_g/c$ of a simulation ($t_f$ stands for the final runtime, see table \ref{tab:parametersSim}). The zoomed-in region highlights the similarity of this fraction between the simulations. Right panel: Time-averages of the fraction shown in the left panel. The error bars correspond to one standard deviation divided by the square root of the number of data points.}
    \label{fig:konstants}
\end{figure*}

The left panel of Fig. \ref{fig:konstants} shows the energy flux of the jet divided by $\flux_{jet}^2\anglB{\Omega_H^2 + n\Omega_H^4 + m\Omega_H^6}$ within the last 5000 $r_g/c$ ($t_f$ stands for the final runtime, see table \ref{tab:parametersSim}). We summed the energy and magnetic fluxes, evaluated at the horizon, of the northern and southern jet regions. For all but the MAD models, the ratio stays constant. The right panel shows time-averages of the ratios from the left panel. 
The cyan dotted line marks the expected value of $k$ for a split-monopole geometry. Visibly, the ratios corresponding to $k$ are close to, but below, the split-monopole value. The constants of the $\beta=1$ (grey error bar) and $\beta=5$ (blue error bar) simulations deviate with $k\approx0.49$ the most, but these two simulations are also heavily affected by the magnetic flux eruptions. For the other simulations, the ratios differ from $\anglB{6\pi}^{-1}$ by less than 6 percent. We therefore conclude that the power extraction from the black hole can well be described by the BZ process of a magnetosphere that is close to monopolar near the horizon.  

\begin{figure*}
    \centering
    \includegraphics[width=0.9\linewidth]{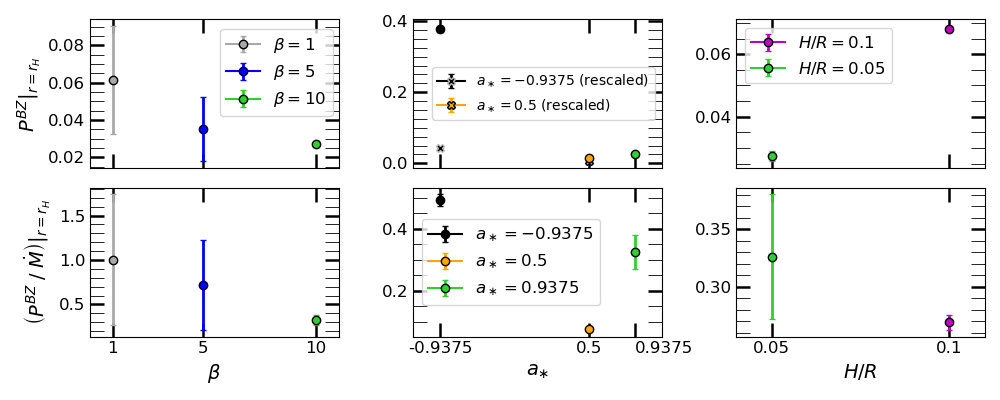}
    \caption{Top row: Jet power as measured at the horizon and time-averaged over the last respective $5000r_g/c$. Bottom row: Same as the top row, but showing the jet efficiencies. The labels apply to one column respectively. To account for the spin-dependence in the initial conditions (see section \ref{subsec:runSim}), we also show the jet power of the $a_*=0.5$ and $a_* = -0.9375$ simulations divided by $\rescale$ (cross markers). Meeting theoretical expectations, the jet is more efficient in more magnetized settings and for more rapidly spinning black holes.}
    \label{fig:jetPowerEff}
\end{figure*}

The time-averaged jet powers measured at the horizon are shown in the top row of Fig. \ref{fig:jetPowerEff}. For the $a_*=0.5$ and $a_* = -0.9375$ simulations (second column), we show the unscaled jet powers (circle markers) and the jet powers divided by $\rescale$ (cross markers). The MAD simulations (grey errorbar for $\beta=1$ and blue errorbar for $\beta=5$, first column) differentiate themselves from the SANE setups with their relatively large errorbars. Hence, the jet powers of the $\beta=1$ and the $\beta=5$ simulations agree with $P^{BZ}|_{r=r_H}\approx\anglB{6.2\pm2.9}\times10^{-2}$ and $P^{BZ}|_{r=r_H}\approx\anglB{3.5\pm1.7}\times10^{-2}$, respectively, with the fiducial one (green errorbars, $P^{BZ}|_{r=r_H}\approx\anglB{2.7\pm0.2}\times10^{-2}$) due to their large uncertainties. A comparison with the right panel of Fig. \ref{fig:phiTrends} shows that the jet powers follow the same trends as $\flux$ measured at the horizon, except when the black hole spin varies (top row, middle panel): Despite the $a_*=0.5$ simulation accumulating more flux at the horizon, its jet power is with $P^{BZ}\approx1.6\times10^{-2}$ (orange errorbar) less than for the fiducial case. 

The bottom row of Fig.\ref{fig:jetPowerEff} shows the time-averaged jet efficiencies, $P^{BZ}/\mdot$, evaluated at the horizon. The efficiency trend with $\beta$ matches that of the jet power (leftmost column); the efficiencies of the $\beta=1$ and $\beta=5$ simulations in particular are consistent with $100\%$ energy extraction from accreting matter. However, due to the significant variability, all jet efficiencies agree with each other within their uncertainties.

The differences in time-averaged jet efficiencies are more pronounced as a function of spin (middle panel). The jet in the $a_*=0.5$ simulation (orange errorbar) is the least efficient; the ratio of its efficiency to that of the jet in the fiducial simulation is $\approx4.2$. The jet efficiency is more enhanced in the counterrotating simulation (black errorbar): its time-averaged jet efficiency is $\approx0.49$, which is $\approx1.5$ times greater than for the fiducial simulation. This result differs from ideal GRMHD simulations of thick accretion flows, where the counterrotating setups resulted in less efficient jets \citep[e.g.,][]{Tchekhovskoy2012MNRAS.423L..55T,Narayan2022MNRAS, Dhang2025ApJ}. 
For the jet efficiency, dividing by the mass accretion rate is supposed to account for more magnetic flux dragged towards the black hole, which enhances $P^{BZ}$. 

\begin{figure*}
    \centering
    \includegraphics[width=0.9\linewidth]{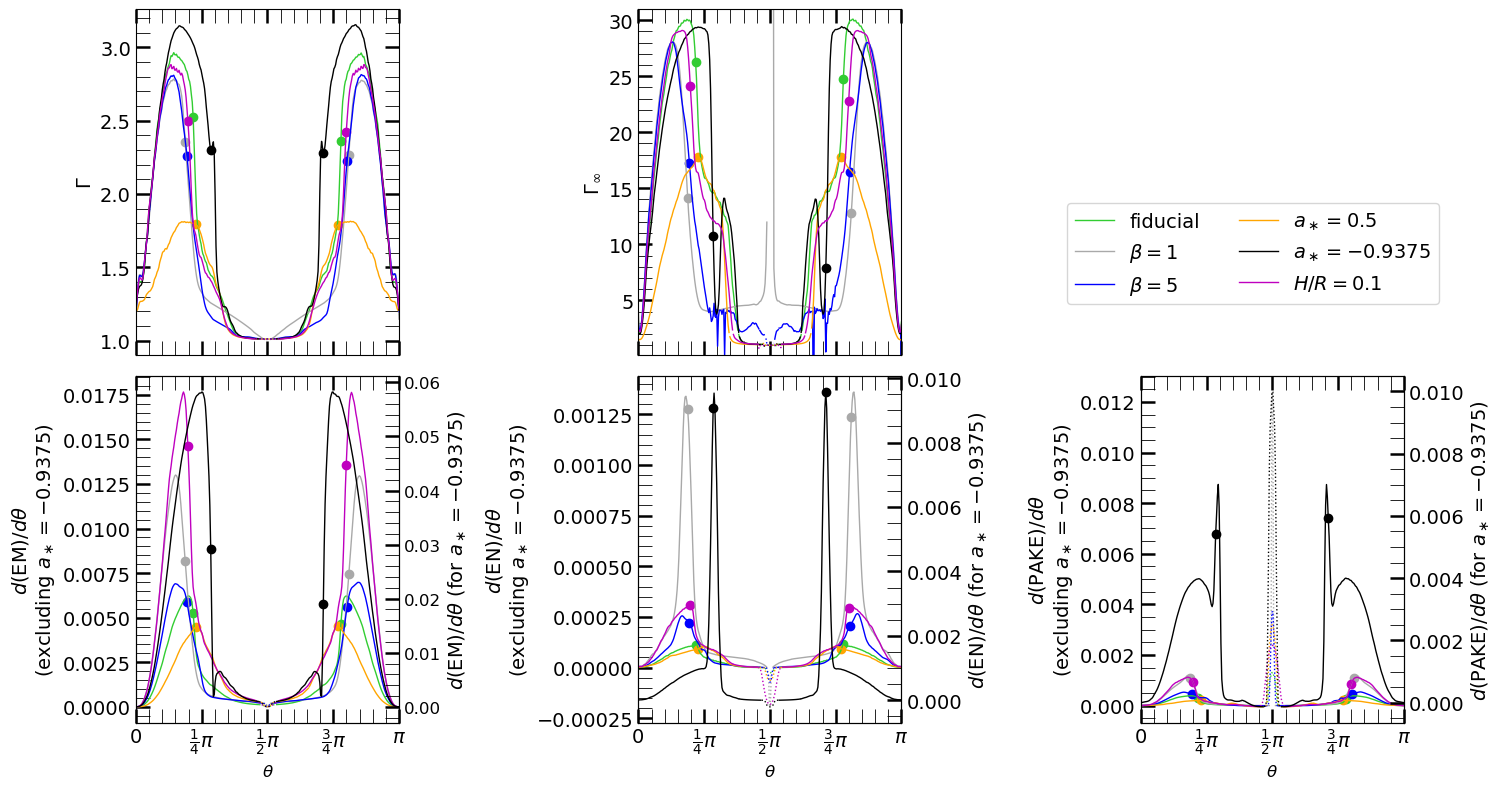}
    \caption{Angular profiles of accretion properties at $r=50r_g$, time-averaged over the respective last $5000r_g/c$. Lines switch from solid to dotted within the disk region (where $Be \leq 0$, eq. \ref{eq:bern}). Scatter points mark the location of the jet boundary (right side of Fig. \ref{fig:sigmaFieldLineContour}). Top row: Lorentz factor for an Eulerian observer, $\Gamma=\anglB{1-v_iv^i}^{-1/2}$ (left panel), and $\Gamma$ at infinity (right panel, eq. \ref{eq:gammaInf}). Negative and very large $\Gamma_\infty$ visible in the profiles of the $\beta=1$ (grey line) and $\beta=5$ simulations result from sign changes in $\mdot$. Bottom row: Same as the top row, but showing the EM (left panel), EN (middle panel) and the PAKE stress energy component (right panel). We utilize the right y-axis for the $a_* = -0.9375$ simulation (black lines), because its stress-energy tensor components significantly exceed those of the other simulations. At $\approx0.3\pi$ and $\approx0.7\pi$, $B^\phi$ changes sign in $a_* = -0.9375$ simulation, enhancing the EN and PAKE components while decreasing the EM term.}
    \label{fig:slices50}
\end{figure*}

To examine the angular variations of the jet properties, we create a time-averaged slice at $50r_g$ (Fig. \ref{fig:slices50}). The scatter points mark the location of the jet boundary (right side in Fig. \ref{fig:sigmaFieldLineContour}), and dotted lines show the properties in the disk region (where $Be\leq0$, eq.\ref{eq:bern}). Unsurprisingly, matter moves the slowest in the disk with the Lorentz factor for an Eulerian observer $\Gamma=\anglB{1-v_iv^i}^{-1/2}\approx1$ (left panel in top row). In the wind and jet region, $\Gamma$ reaches $\gtrsim 1.8$ for all simulations and is the smallest for the $a_*=0.5$ simulation (orange line). The angular profiles of the Lorentz factors of the $\beta=1$ (gray line) and $\beta=5$ (blue line) simulations are similar, and both reach a maximum of $\Gamma\approx2.8$ at $50 r_g$. Their smaller Lorentz factors, compared with the fiducial (green line), the $H/R=0.1$ (purple line), and the $a_*=-0.9375$ (black line) simulations, might be due to time-averaging over highly variable simulations and to the dissipated magnetic energy not yet having been converted into particle acceleration.

As visible in the top-right panel, the jets in all simulations reach high Lorentz factors at infinity, assuming all energy is converted into bulk kinetic motion (eq. \ref{eq:gammaInf}). Because of density floors (section \ref{subsec:numset}), the plotted $\Gamma_\infty$ represent a lower limit. Magnetic reconnection present in the $\beta=1$ (grey line) and $\beta = 5$ (blue line) simulations leads to varying in- and outflows, i.e., sign changes in $\mdot$, resulting in their shown unphysically high or negative $\Gamma_\infty$. Except for the weak jet in the $a_*=0.5$ simulation (orange line), whose time-averaged $\Gamma_\infty$ peaks at $\approx18$, all simulations have $\Gamma_\infty$ maxima $\gtrsim28$. Note that the wind region (right/left of the scatter points of the northern/southern region) can reach high Lorentz factors as well. We will examine the wind velocities in more detail in section \ref{subsec:windV}.




Because the jet is driven by Poynting flux, the EM component (left panel in bottom row) dominates the EN and PAKE components (middle and right panels in bottom row). For this row, we use the right y-axis to plot the stress-energy components of the $a_* = -0.9375$ simulation (black lines), because they far exceed those of the other simulations, likely due to the larger overall density of the initial conditions for the counter-rotating case (see section \ref{subsec:runSim}). For all simulations, the maximum of the time-averaged EM component in the northern unbound region, i.e. $\theta < \pi/2$, is $\gtrsim8$ times greater than the maximum PAKE component and $\gtrsim6$ times greater than the maximum EN component. 
The ratio between the PAKE and the EM component is with $\approx22.6$ the greatest for the $a_*=0.5$ simulation (orange lines). The EN-to-EM ratio is maximized with $\approx56.2$ for the $H/R=0.1$ simulation (purple lines). To compare maxima of the stress-energy components among simulations, the jet width and time-variations in the jet profile need to be considered. Because the jet energy flux is conserved, a larger jet opening angle spreads the jet energy over a larger solid angle, reducing the respective component maximum. Also, since Fig.\ref{fig:slices50} depicts time-averaged properties, slight angular movements of the jet will lower the time-averaged maximum at a specific angle. Thus, 

For the counterrotating simulation, the time-averaged EN and PAKE components peak at distinctively different angles compared to the EM component. The EN and PAKE components spike at $\approx52^\circ$ and $\approx127^\circ$, while the EM component drops. At these angles, $B^\phi$ changes sign in a current sheet (right panel in bottom row of Fig.\ref{fig:theEndRho}), decreasing the EM term. Dissipation in the current sheet causes heating (EN) and bulk acceleration (PAKE). The Lorentz factor at infinity decreases with increasing density (eq.\ref{eq:gammaInf}). Hence, the matter outflow in the current sheet decreases $\Gamma_\infty$. As we will show in section \ref{subsec:windV} and examine its causes in more detail in section \ref{subsub:lackmassloading}, appreciable mass outflow in the highly magnetized wind region only occurs in the $a_*=-0.9375$ simulation, and only in the current sheet.

\subsection{BP outflow}\label{subsec:windprop}
\subsubsection{Wind velocities and mass loading}\label{subsec:windV}
\begin{figure*}
    \centering
    \includegraphics[width=0.9\linewidth]{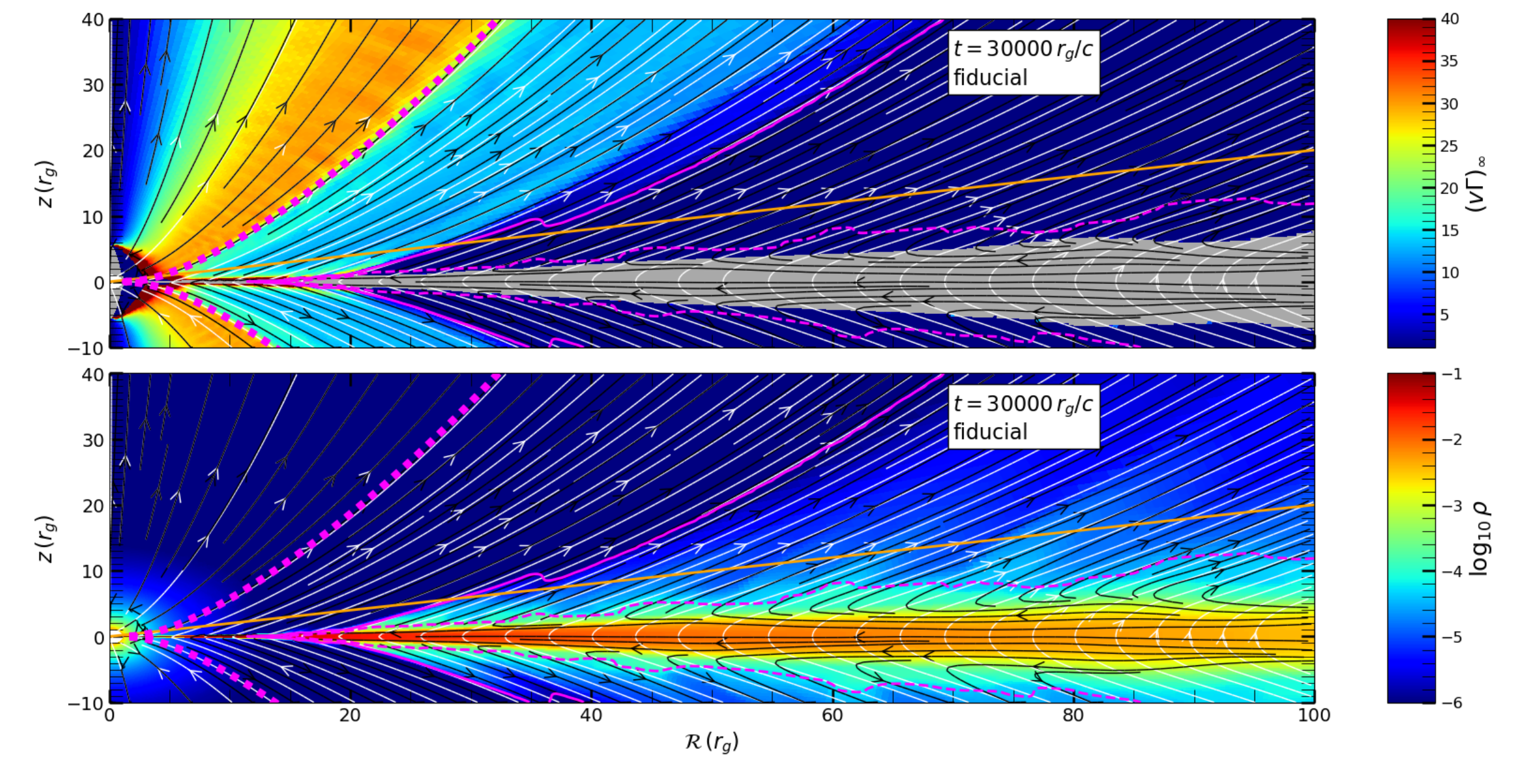}
    \caption{Top panel: Map of $\anglB{v\Gamma}_\infty$ of the fiducial simulation at $t=30000r_g/c$. As in Fig.\ref{fig:theEndRho}, the white solid line marks the jet boundary, the magenta solid line follows the $\sigma=1$ contour line, and the magenta dashed line traces $Be=0$. The black streamlines represent the velocities and align with the magnetic field lines (white arrows) at large radii. The values plotted in Fig. \ref{fig:outFlowsVelocity} were extracted along the orange solid line. Bottom panel: Same as the top panel, but showing the density.}
    \label{fig:windWedgeRegions}
\end{figure*}

Fig. \ref{fig:windWedgeRegions} shows $\anglB{v\Gamma}_\infty$ (top panel) and the logarithm of the density (bottom panel) for the fiducial simulation at its last time step. The magenta lines are the same as in Fig. \ref{fig:theEndRho}, referring to the jet boundary (dotted lines), the $\sigma=1$ contour (solid lines), and the $Be=0$ contour (eq.\ref{eq:bern}, dashed lines). Here, the magnetic field follows the white streamlines. As already indicated by Fig. \ref{fig:slices50}, high velocities (black streamlines) are not exclusive to the jet region but also appear in the highly magnetized wind region, with $\anglB{v\Gamma}_\infty>10$. However, the bottom panel does not indicate any visible mass loading in this highly magnetized region; winds seemingly remove disk material only where $\sigma<1$.

\begin{figure*}
    \centering
    \includegraphics[width=0.9\linewidth]{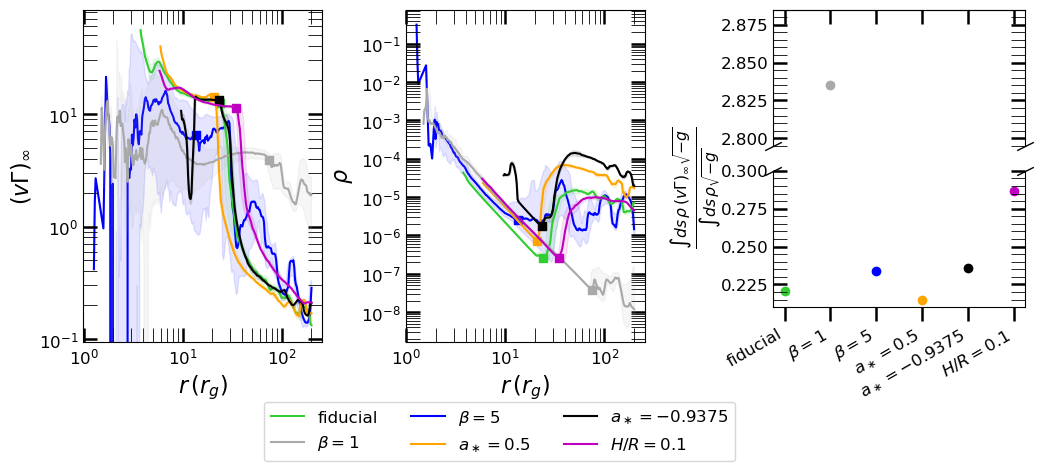}
    \caption{Left panel: Radial profiles of $\anglB{v\Gamma}_\infty$ along an angle of $4H/R$ as measured from the equatorial plane in the wind region (see also top panel of Fig. \ref{fig:windWedgeRegions}). The solid lines correspond to the median values of the last $5000r_g/c$, and the shaded regions span the first and last quartiles. The square markers show the radii up to which density floors are present. Middle panel: Same as the right panel, but showing the densities (see also bottom panel of Fig. \ref{fig:windWedgeRegions}). The density floors are visible through the power-law declines. Right panel: Density-weighted integrals of $\anglB{v\Gamma}_\infty$. For these integrals, we use the median values presented in the left and middle panels where $\anglB{v\Gamma}_\infty\geq1$, and integrate up to $200r_g$ ($s$ is the internal radial coordinate, eq. \ref{eq:intCorS}). As the integral gives more weight to larger radii, the integration results are less affected by density floors.}
    \label{fig:outFlowsVelocity}
\end{figure*}

To analyze the wind launching, we select an angle of $4H/R$ as measured from the equatorial plane (orange solid lines in Fig. \ref{fig:windWedgeRegions}). Along this angular slice, we plot the radial profiles of the product of the speed and the Lorentz factor at infinity (left panel in Fig. \ref{fig:outFlowsVelocity}, see also the top panel in Fig. \ref{fig:windWedgeRegions}), the densities (middle panel in Fig. \ref{fig:outFlowsVelocity}, see also the bottom panel in Fig. \ref{fig:windWedgeRegions}) and the density-weighted integral of $\anglB{v\Gamma}_\infty$ (right panel) in the wind region (i.e., unbound material beyond the contours shown in the right side of Fig. \ref{fig:sigmaFieldLineContour}). For the radial profiles and the integration, we use the median values within the respective last $5000r_g/c$ (the shaded regions of the radial profiles correspond to the first and last quartiles). For the integration (right panel), we additionally require that $\Gamma_\infty\geq 1$. This requirement, alongside taking the median, filters out non-steady dynamics in the $\beta=1$ (gray lines and marker) and $\beta = 5$ (blue lines and marker) simulations, as already discussed for Fig. \ref{fig:slices50}.

As already visualized in Fig.\ref{fig:windWedgeRegions} for the fiducial simulation, the velocities of the simulated disk winds are the greatest at small radii where $\sigma > 1$ (left panel of Fig. \ref{fig:outFlowsVelocity}). The median radial profiles of $\anglB{v\Gamma}_\infty$ at an angle $4H/R$ reach $\gtrsim11$ for the fiducial (green line), $a_* = 0.5$ (orange line) and $H/R = 0.1$ (purple line) simulations. The $\anglB{v\Gamma}_\infty$ briefly drops to $\approx1.8$ at $r\approx12r_g$ due to the present current sheet for the $a_* = -0.9375$ simulation (black line). The velocities of the $\beta = 1$ (gray line) and $\beta = 5$ (blue line) simulations are smaller; within $10r_g$, the average of their medians is $\lesssim7$. Their low $\anglB{v\Gamma}_\infty$ profiles stem from their non-steady dynamics, yielding even negative median $\anglB{v\Gamma}_\infty$ values for the $\beta = 5$ simulation.

The density profiles (middle panel) of the $\beta = 1$ (gray line) and $\beta = 5$ (blue line) simulations are more comparable to those of the other ones, decreasing at small radii before increasing again. All median profiles in the wind region feature the same power-law dependence along some radial interval. This dependence is caused by our density floors (see section \ref{subsec:numset}) getting activated in the $\sigma>1$ wind. The density map plotted in the bottom panel of Fig. \ref{fig:windWedgeRegions} for the fiducial simulation at $30000r_g/c$ visualizes that the power-law density profiles are not connected to mass outflows. The wind region dominated by our floor model is indicated by square scatter points, marking where the power-law dependence of the densities ends, in the left and middle panels of Fig. \ref{fig:outFlowsVelocity}. For a lower floor model, the actual $\anglB{v\Gamma}_\infty$ values are expected to be even higher. The density floors not only affect $\anglB{v\Gamma}_\infty$ but also the measured mass loading at small radii. Further analysis of the mass loading that accounts for the density floors can be found in appendix \ref{ap:massloadwind}. We examine the lack of mass outflows from the inner disk region in section \ref{subsub:lackmassloading}.  

The mass-weighted $\anglB{v\Gamma}_\infty$ are less affected by density floors (right panel) due to the weighting with the volume element. 
All integrals are evaluated along the 4H/R line up to 200 $r_g$, and $s$ refers to the internal radial coordinate (eq.\ref{eq:intCorS}). The $\beta=1$ simulation (gray dot) has, with an integration result of $\approx2.84$, the largest potential to launch relativistic disk winds. However, it is also the one where the disk-connected field lines are almost entirely loaded with floor material.  
The winds in the other simulations reach integrated $\anglB{v\Gamma}_\infty$ ranging from $\approx0.21$ for the $a_* = 0.5$ simulation (orange dot) to $\approx0.29$ for the $H/R=0.1$ one (purple dot). A close inspection shows that ranking the integrated $\anglB{v\Gamma}_\infty$ values is similar to ranking the jet powers (top row of Fig. \ref{fig:jetPowerEff}): a simulation with a strong jet also has a fast disk wind. This feature is connected to the wind power scaling in tandem with the jet power, as we will show in section \ref{subsec:windPower}.


\subsubsection{Wind Power}\label{subsec:windPower}

\begin{figure*}
    \centering
    \includegraphics[width=0.9\linewidth]{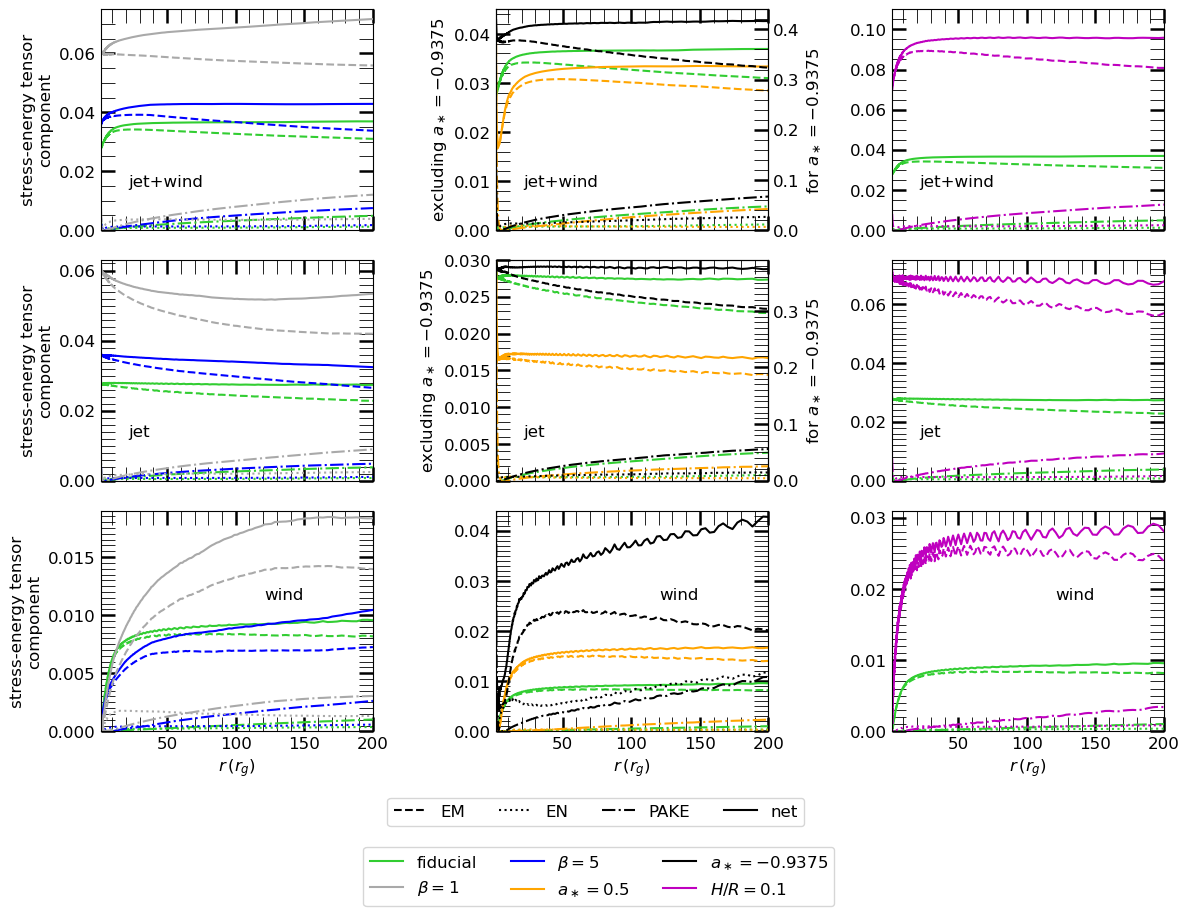}
    \caption{Radial profiles of time-averaged wind and jet powers. The figure shows the radial profiles of the net energy flux (solid lines), decomposed into its electromagnetic (dashed lines, eq. \ref{eq:Tem}), free-particle (dot-dashed lines, eq. \ref{eq:Tpake}), and enthalpy (dotted lines, eq.\ref{eq:Ten}) components. For the profiles, we summed the northern and southern components and time-averaged them over the last $5000r_g/c$. Note that for the first two panels of the middle column, the left y-axis is for the fiducial (green lines) and the $a_\ast=0.5$ simulation (orange lines), while the right y-axis is for the $a_\ast=-0.9375$ simulation (black lines). The constant net energy flux (solid lines) of the combined wind and jet (top row) shows that energy is conserved. For all simulations except the $a_\ast=0.5$ one, the jet (middle row) is more powerful than the disk wind (bottom row). These outflows are initially dominated by the electromagnetic component, which is converted to heat (EN) and bulk acceleration (PAKE) at larger radii. 
    }
    \label{fig:jetWind}
\end{figure*}
Fig. \ref{fig:jetWind} shows radial profiles of the stress-energy tensor components of the jet and wind combined (top row), the jet (second row), and the wind (bottom row), time-averaged over the respective last $5000r_g/c$. For the first and second panels of the middle column, the left axis is for the fiducial (green lines) and the $a_*=0.5$ (orange lines) simulations, while the right axis is for the counterrotating simulation (black lines). The net energy flux of the combined outflow still increases at large radii for the $\beta=1$ simulation (grey line in left panel), which indicates that its energetics could be impacted by floors. Otherwise, the net energy flux (solid lines) of the combined outflows show good energy conservation, varying by less than 1 percent between $150r_g$ and $200r_g$. 

We separate the net outflow energetics as belonging to the jet (second row) or the wind (bottom row) using the last magnetic field line anchored at the black hole (using eq.\ref{eq:aphiTime}, right side of Fig. \ref{fig:sigmaFieldLineContour}). The wiggles in the stress-energy component profiles arise from interpolating the simulation data onto a discrete grid to obtain the jet boundaries. Due to the wiggles, we average the components between $150r_g\leq r \leq 200 r_g$ for any comparison. With the chosen jet boundaries, the net energy fluxes of the jet (solid lines, middle row) at large radii agree well with the jet powers measured at the horizon (top row of Fig. \ref{fig:jetPowerEff}) for the SANE simulations.


All simulated winds are Poynting-flux dominated (bottom row). The EM component (dashed lines) contributes the least to the net energy flux in the $a_*=-0.9375$ simulation (black lines in the middle column) compared with the other simulations, due to the current sheet (bottom row of Fig. \ref{fig:theEndRho}). Still, for that simulation the time-averaged EM component (eq.\ref{eq:Tem}) is $\approx1.9$ times and $\approx2.2$ times greater than the EN (eq.\ref{eq:Ten}) and the PAKE component (eq.\ref{eq:Tpake}) as averaged between $150r_g\leq r\leq 200r_g$, respectively. In the MAD simulations, magnetic reconnection converts Poynting flux to heat (EN) and bulk acceleration (PAKE). The sum of the EN and PAKE components makes up $\approx23$ percent ($\beta=1$) and $\approx29$ percent ($\beta=5$) of the total time-averaged wind power. For the SANE simulations, the EN and PAKE terms contribute $\lesssim14$ percent of the total wind power.

The jet powers averaged between $150r_g\leq r\leq 200r_g$ slightly differ with those measured at the horizons. The $a_*=0.5$ simulation (orange lines, middle column) has with $4$ percent the greatest discrepancy among the SANE simulations, while the time-averaged jet powers differ by $\lesssim1$ percent as compared to $P^{BZ}$ measured at the horizon for the other SANE simulations. For the MAD simulations, the difference between the time-averaged jet power between $150r_g\leq r\leq 200r_g$ and at the horizon is $\approx14$ percent (for $\beta=1$) and $\approx6$ percent (for $\beta=5$). Besides density floors, this difference also stems from limitations in the resolution, which allow the jet energy flux to extend beyond the outermost magnetic field line anchored at the black hole. However, as the comparison of jet powers shows, our calculated outermost magnetic field lines are adequate tracers of the jet boundary.

\begin{figure*}
    \centering
    \includegraphics[width=0.9\linewidth]{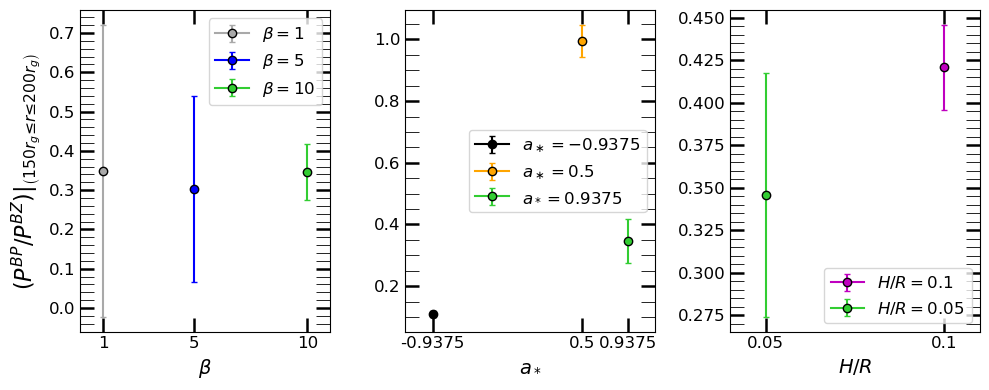}
    \caption{Wind versus jet power. The error bars show the ratios of the time-averaged net energy fluxes (solid lines in Fig. \ref{fig:jetWind}), averaged between $150r_g\leq r\leq 200 r_g$. The wind power relative to the jet is marginally affected by a change in $\beta$ (left panel). The wind power relative to the jet increases with a smaller (prograde) black hole spin (middle panel) or with an increase in the disk thickness (right panel).}
    \label{fig:jetWindComparison}
\end{figure*}

Fig. \ref{fig:jetWindComparison} shows the ratio of the time-averaged wind and jet powers between $150r_g\leq r \leq 200 r_g$ (see also table \ref{tab:jetwind}). The effect of a stronger jet with a greater magnetic field (i.e., smaller $\beta$, first column in Fig. \ref{fig:jetWind}) is canceled out by an enhanced wind power, such that the wind versus jet power is approximately constant as $\beta$ varies (leftmost panel in Fig. \ref{fig:jetWindComparison}); the wind is always $\approx3$ times weaker than the jet. 

Changing the black hole spin has the greatest impact on the jet versus wind power (middle panel in Fig.\ref{fig:jetWindComparison}, middle column in Fig.\ref{fig:jetWind}). While both the rescaled wind and jet power are greater for the $a_* = -0.9375$ simulation than for the fiducial one, the retrograde setup produces the weakest wind compared to its jet; the wind is about $9$ times weaker than its jet if $a_* = -0.9375$. The relatively weak wind of the $a_* = -0.9375$ might be connected to the broadness of its jet, which reduces the area of the highly magnetized wind region (also visible when comparing the radial ranges left to the squares in the left and middle panels of Fig.\ref{fig:outFlowsVelocity}). The $a_* = 0.5$ simulation (orange error bar in middle panel) launches a wind approximately as strong as its jet. 
As visible in table \ref{tab:jetwind}, the wind power varies less than the jet power as a function of spin, so that the middle panel of Fig. \ref{fig:jetWindComparison} is primarily driven by variations in the BZ power.

Changing the disk thickness changes the wind and jet power to the same extent (right panel of Fig.\ref{fig:jetWindComparison}). The jet power of the $H/R=0.1$ simulation is $\approx\anglB{3\pm 0.6}$ times greater than the fiducial simulation, while the wind power is $\approx\anglB{2.5\pm 0.2}$ times greater (see also the rightmost column of Fig. \ref{fig:jetWind}). Consequently, the ratios of wind to jet power of the fiducial and the $H/R=0.1$ simulations (rightmost panel in Fig. \ref{fig:jetWindComparison}) agree with each other within their uncertainties (see also table \ref{tab:jetwind}). Hence, the strength of the wind versus the jet hardly depends on disk thickness, leaving the black hole spin (middle panel) as the strongest influencer of this ratio.

\begin{table*}
\centering
\begin{tabular}{|c|c|c|c|c|c|c|} 
\hline
\begin{tabular}[c]{@{}c@{}}Power \\between \\$150r_g\leq r\leq200 r_g$\end{tabular} & fiducial             & $\beta = 1$ & $\beta = 5$ & $a_* = 0.5$ & $a_* = -0.9375$ & $H/R = 0.1$  \\ 
\hhline{|=======|}
$P^{BZ}\,\anglB{10^{-2}}$  & $2.7\pm0.2 $  & $5.3\pm1.8 $  &   $3.3\pm1.7 $ & $1.7 \pm 0.1$ & $37.5\pm1.0 $ &$6.7\pm0.2 $ \\ 
$P^{BZ}\,\anglB{10^{-2}}$ (rescaled)&  &    & & $0.6\pm0.0$ & $4.2\pm0.1 $ & \\
\hline
$P^{BP}\,\anglB{10^{-2}}$& $0.9\pm0.2 $ & $1.8\pm1.9$   &$1.0\pm0.6 $& $1.7\pm0.1$ & $4.1\pm0.3 $ &$2.8\pm0.2 $ \\
$P^{BP}\,\anglB{10^{-2}}$ (rescaled)&  &    & & $0.6\pm0.0$ & $0.5\pm0.0 $ & \\
\hline
$P^{BP}/P^{BZ}\,\anglB{10^{-2}}$& $34.6\pm7.2 $ & $34.9\pm37.2 $   &$30.3\pm23.7 $&   $99.5\pm 5.3$& $11.0\pm1.0$ &$42.1\pm2.5 $ \\
\hline
\end{tabular}\label{tab:jetwind}\caption{Time-averaged jet (top two rows) and wind (middle two rows) power, averaged between $150r_g\leq r\leq200 r_g$ (solid lines in second and third row of Fig. \ref{fig:jetWind}). For the $a_*=0.5$ and $a_*=-0.9375$ simulations, we also show the jet and wind powers divided by $\rescale$. The ratios between the wind and jet powers (bottom row) are plotted in Fig.\ref{fig:jetWindComparison}.}
\end{table*}

\section{Discussion}\label{sec:discussion}

\subsection{Presence of a high-velocity wind and the lack of mass loading}\label{sec:whyWind}
\subsubsection{Lack of mass loading}\label{subsub:lackmassloading}
Our simulations feature a fast wind with $\anglB{v\Gamma}_\infty > 10$ in the $\sigma>1$ region (left panel of Fig. \ref{fig:outFlowsVelocity}). However, this fast wind does not carry any measurable amount of matter; the densities in the fast-wind region are dominated by density floors instead (middle column of Fig. \ref{fig:outFlowsVelocity}). 

\begin{figure*}
    \centering
    \includegraphics[width=0.7\linewidth]{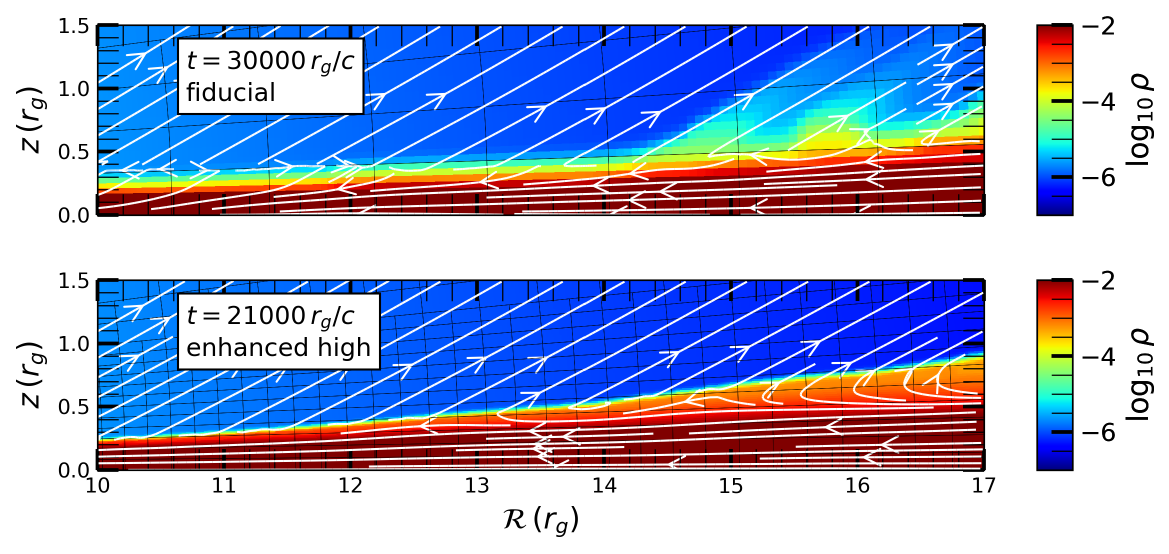}
    \caption{Close-up of the density map in the fast-wind region of the fast-wind region of the fiducial (top panel) and the high-resolution simulations with four AMR levels (bottom panel). We refer to the enhanced high-resolution simulation as "enhanced high" instead of its signifier "high" (see table \ref{tab:parametersSim}) to highlight the additional AMR level on top of the "high" simulation. The velocity trajectories (following the white arrows) do not differ significantly between the two simulations despite the different resolutions (the AMR blocks are represented by the black grid).}
    \label{fig:zoomInFidDubDub}
\end{figure*}

The lack of mass loading in this region could be due to an insufficient resolution not accurately capturing forces for wind-launching. To test this, we examine the wind-launching conditions in the simulation with twice the resolution (called "high" in table \ref{tab:parametersSim}). Zooming into the fast-wind region, we see no noticeable difference in the velocity streamlines between the simulations with the naked eye. Moreover, the material is still floored in the "high" simulation. Hence, we add another AMR level on top of the high-resolution simulation and let the simulation evolve for an additional $1000r_g/c$. Fig. \ref{fig:zoomInFidDubDub} shows snapshots of the density maps for the fiducial (top panel) and the enhanced high-resolution simulations (bottom panel) in the fast-wind region. The black grid lines represent the AMR blocks. At radii $r\gtrsim14r_g$, the disk in the enhanced high-resolution simulation is thicker, but this may be due to that simulation snapshot being from an earlier timestep. Despite slight differences in the density profiles, the outflow behavior is visibly very comparable (white arrows follow velocity streamlines), and in neither simulation, an appreciable amount of matter is removed from the disk in this region. Thus, increasing the resolution four-fold to an effective $N_{\theta} = 3072$ cells does not lead to any measurable difference in the mass-loading of the highly magnetized region. Either the required resolution would need to be a multitude larger, or the lack of wind launching is due to a resolved physical effect.  

\begin{figure*}
    \centering
    \includegraphics[width=0.9\linewidth]{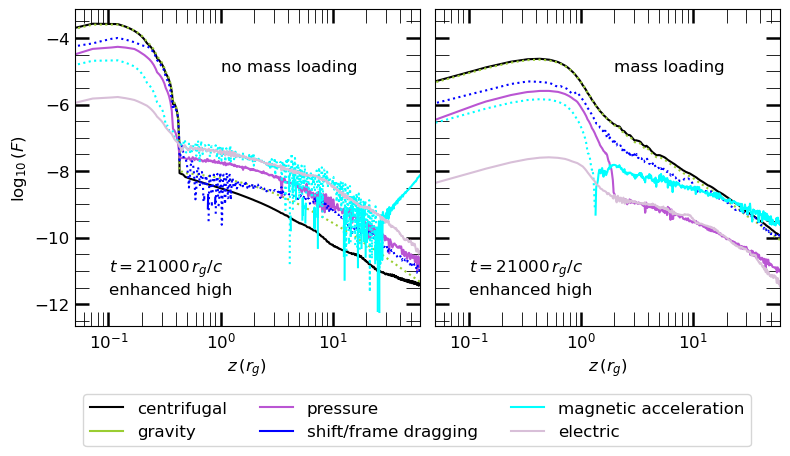}
    \caption{Left panel: Forces along a poloidal magnetic field line tracing $\Tilde{A}_\flux = 4$, where no mass loading occurs. The forces were calculated for the "enhanced high" simulation (i.e., the "high" simulation with an additional AMR level) at $t=21000r_g/c$. Forces are negative where their respective lines are dotted. Right panel: Same as the left panel, but for the poloidal magnetic field line tracing $\Tilde{A}_\flux = 5.5$. A centrifugally driven wind (black line) successfully removes matter from the disk.}
    \label{fig:forcesLines}
\end{figure*}

Assuming the latter, we inspect the forces along poloidal magnetic field lines. We select one magnetic field line, tracing $\Tilde{A}_\flux = 4$, where no mass-loading occurs, and another one, tracing $\Tilde{A}_\flux = 5.5$, that traverses the mass-loaded wind region. Generalizing the force decomposition of \cite{Porth2010ApJ...709.1100P} to the general relativistic case, for the "enhanced high" simulation at $t=21000r_g/c$, we calculate the centrifugal, gravitational, pressure, shift/frame dragging, magnetic acceleration, and electric forces as seen by a Kerr-Schild Eulerian observer (see Appendix \ref{ap:forceBalance} for the derivation of these terms). Note that the electric force projected along the poloidal magnetic field lines is non-zero because of this chosen frame. The left panel in Fig.\ref{fig:forcesLines} shows the forces along a magnetic field line associated with no mass loading. Negative forces are plotted with dotted lines. The centrifugal (black line), pressure (violet line), and electric (light purple line) forces are positive along the entire plotted interval and support wind launching. However, these forces significantly weaken outside of the disk at $z\approx0.4r_g$. Thus, gravity (green line), shift/frame dragging (blue line), and magnetic compression (cyan line) can successfully prevent wind launching. The magnetic acceleration force eventually becomes positive and accelerates the plasma, but only at $z\simeq 30 r_g$ above the disk.

The right panel of Fig.\ref{fig:forcesLines} shows the forces along the poloidal field line tracing $\Tilde{A}_\flux = 5.5$. Along this field line, the centrifugal force, essential for the BP mechanism, aided by pressure and the electric force, successfully lifts disk material. In contrast to the no-mass loading case (right panel), the magnetic acceleration force is negative inside the disk but switches sign at the disk surface, which has been identified as a feature of BP-driven winds by e.g., \cite{Ferreira1997A&A...319..340F}. Thus, Fig. \ref{fig:forcesLines} not only depicts the forces that enable or prevent mass loading, but also confirms that the BP process drives the winds in our simulations. 

\begin{figure*}
    \centering
    \includegraphics[width=0.9\linewidth]{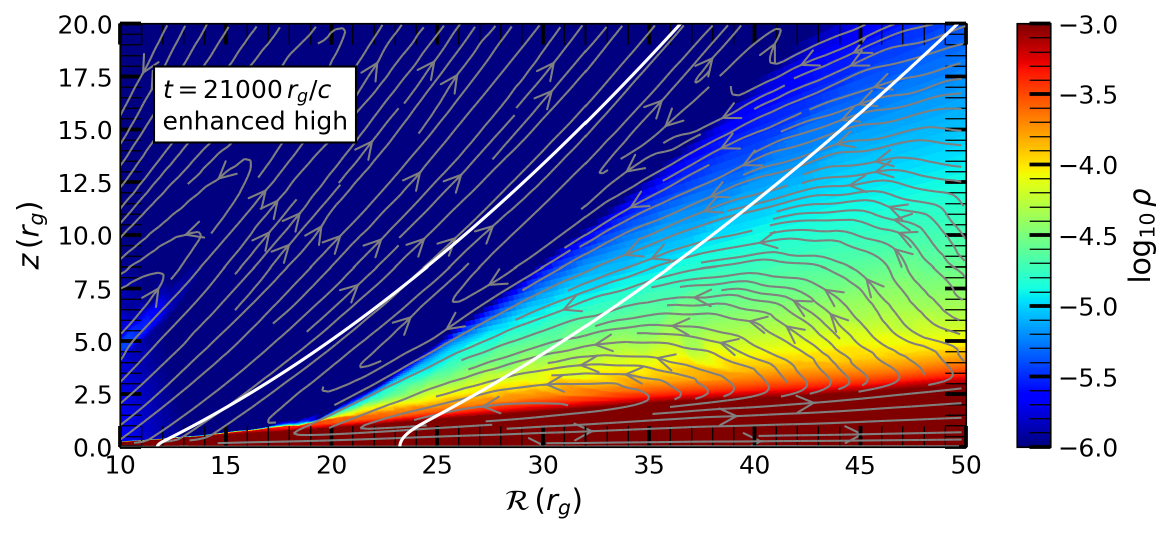}
    \caption{Density map of the "enhanced high" simulation (i.e., the "high" simulation with an additional AMR level) at $t=21000r_g/c$. The forces projected along the plotted poloidal field lines (white solid lines) are featured in Fig. \ref{fig:forcesLines}. The grey streamlines follow the currents.}
    \label{fig:currents}
\end{figure*}

The Lorentz force providing magnetic acceleration (cyan lines in Fig.\ref{fig:forcesLines}) and either supporting or preventing mass loading arises from the topology of the electric currents in our simulations. Fig.\ref{fig:currents} shows a snapshot of the "enhanced high" simulation at $t=21000r_g/c$ with the magnetic field lines used for Fig.\ref{fig:forcesLines} (white lines). The currents (grey streamlines) follow two loops, one inside the fast-wind region with no mass loading, and one where mass loading occurs. The crossproduct of the current and the magnetic field gives rise to the Lorentz force. Where no mass is loaded onto the field lines, the Lorentz force compresses the disk and remains negative for several $r_g$ above the disk. Where mass loading occurs, the Lorentz force is negative within the disk as well, but switches sign at the disk surface. Hence, the current loop structure in our simulations significantly affects the magnetic field's ability to carry away disk material.   
\subsubsection{Observational implications of a high-velocity wind}\label{subsub:obshighwind}
The fast wind in the highly magnetized region could be considered part of the jet. Since the fast wind transports hardly any matter, it will not contribute to observed emission or absorption lines. Thus, the fast wind and the jet may be difficult to distinguish in observations. \cite{Xie2012RAA....12..817X} proposed a two-component jet model: the first component is a BZ-driven jet, and the BP mechanism launches the second component. For a spin of $a_*=0.9375$ and setting the parameters of their model $\alpha=3$ and $s =0.01$, their analytical functions yield a wind Lorentz factor of $\approx32$ at $r=r_{ms}$ (eq.\ref{eq:rms}), which is comparable to the maximum $\Gamma_\infty$ measured in the wind region (right panel in top row). Their analytical calculations can yield even higher wind Lorentz factors depending on the model parameters. Unfortunately, density floors prevent us from estimating the model parameter values from our simulation for a more thorough comparison with the model by \cite{Xie2012RAA....12..817X}. Nevertheless, our simulations point towards a possible pathway of a two-component jet structure, which has also been proposed to explain the observed luminosity and time evolution of jet measurements of gamma-ray bursts and AGNs \citep[e.g.,][]{Chiaberge2000A&A...358..104C,Filgas2011A&A...526A.113F}. Given the scale-invariance of black hole accretion, our results may apply to stellar-mass black holes in X-ray binaries as well.


\subsection{Simultaneous jet and wind launching}\label{sub:simjetwind}
Our simulated thin accretion disk simulations feature a BZ-driven jet that is always at least as strong as the BP-driven wind down to a black hole spin parameter of $a_*=0.5$ (Fig. \ref{fig:jetWindComparison}). Simultaneous jet and wind launching has been observed in radio-loud AGNs \citep[e.g.,][]{TombesiSambrunaEtAl2010,Giroletti2017A&A...600A..87G}. \cite{TombesiTazakiEtAl2014} estimated that $\approx\anglB{50\pm20}\%$ of their sample of radio-loud AGNs observed featured ultra-fast outflows. \cite{Mehdipour2019A&A...625A..25M} suggested that the relationship between the column density of the ionized wind and the radio-loudness parameter of the jet in their radio-loud AGN sample could be explained if the jet as well as the wind are magnetically driven. While the radio-loudness parameter was inversely related to the wind column density in the sample studied by \cite{Mehdipour2019A&A...625A..25M}, which does not appear to align with our results \citep[note that the alignment will be affected on whether the highly-magnetized wind region is counted towards the jet, following the model by][see also section \ref{subsub:obshighwind}]{Xie2012RAA....12..817X}, our simulations offer a pathway to explaining these observations.

Aligning our simulations with observations of X-ray binaries is not as straightforward. A thin accretion disk may be expected in the spectral state called the soft state \citep[for a review, see e.g.,][]{Remillard2006ARA&A..44...49R}. However, observations found the radio emission of X-ray binaries featuring black holes in the soft state to drop, which has been interpreted as jets launched via the BZ mechanism being strongly quenched \citep[e.g.,][]{Fender2009MNRAS.396.1370F,Russell2011ApJ...739L..19R}. Moreover, \cite{Zhang2026NatAs..10..281Z} observed the black hole X-ray binary 4U 1630-472 to harbor either an X-ray disk wind or a jet, but not both at the same time. Note that wind signatures in, e.g., near-infrared, can persist across spectral states, \citep[e.g.,][]{SanchezSierras2020A&A...640L...3S}, even though their connection to X-ray winds remains unclear \citep{Munoz2022A&A...664A.104M}. Hence, the BZ-powered jet in our thin-disk simulations is not observed in the soft state of X-ray binaries.

However, our simulation setup could correspond a black hole accretion environment in an X-ray binary in the intermediate state. X-ray binaries in the intermediate state display a softer spectrum than in the hard state, along with jet emission \citep[e.g.,][]{Fender2004MNRAS.355.1105F,Russell2013ApJ...768L..35R}. Two intermediate state transitions exist: the first one bridges the hard and the soft state, and the black hole transitions back to the hard state during the second intermediate state. In the state evolution proposed by \cite{Begelman2014ApJ...782L..18B}, a strong magnetic field with $\beta < 100$ strengthens the MRI in the disk during the first intermediate state transition, and the disk changes from a hot thick flow to a thin disk. During the soft state, flux accumulation becomes negligible and flux diffuses outward, increasing $\beta$. The second intermediate state transition starts out with $\beta\gg10^2$, and the \mdotname decreases, which re-enables the formation of a hot thick flow. Given that $\beta$ is considerably smaller in our setups, they may therefore be more comparable to a state between the hard to soft transition rather than vice versa.

\section{Conclusions}\label{sec:conclusions}

Accreting black holes launch magnetically-driven outflows, powered either through the BZ or the BP mechanism. While the resulting jets are a staple of black hole observations \citep[e.g.,][]{Fender2004MNRAS.355.1105F}, their powering mechanism remains uncertain. Understanding the driver of outflows is crucial, though, as black hole outflows have a profound impact on galaxy evolution, black hole growth, and potentially cosmic ray acceleration \citep[e.g.,][]{Tamborra2014JCAP...09..043T,Wang2016NatPh..12.1116W,Bischetti2022Natur.605..244B,Roy2026ApJ..1009L..34R}. But while BZ jets have been extensively studied via GRMHD simulations \citep[e.g.][]{Tchekhovskoy2011MNRAS,PorthChatterjeeEtAl2019a, Liska2020MNRAS,Dhruv2025ApJS}, driving outflows via the BP-mechanism in GRMHD simulations has proven to be challenging.

In this work, we have discussed resistive GRMHD simulations of thin radiatively cooled accretion disks and we use a resistivity profile to model anomalous resistivity originating in (unresolved) turbulent processes. 
We find the collimation of our BZ-driven jets to be comparable to those from ideal GRMHD simulations of MADs, despite the absence of a thick disk. Moreover, their powers closely follows expectations for BZ jet powers  \citep[][section \ref{subsec:jetprop}]{Tchekhovskoy2010ApJ}. The wind regions in the simulations consist of a highly magnetized region with fast winds that only remove negligible amounts of matter from the disk; with a different jet-wind definition, this fast wind could thus also be referred to as a disk-driven jet. In the less magnetized region, a slower wind is launched by the BP mechanism and efficiently carries matter away (section \ref{subsec:windV}). Section \ref{subsub:lackmassloading} shows that the Lorentz force prevents mass loading in the highly magnetized wind region, and proves that the BP mechanism indeed drives outflows in the less magnetized region. In our simulations, the BP powers are weaker than the BZ powers except when the black hole is only slowly spinning, and the BZ-to-BP power ratio is mainly sensitive to black hole spin even after rescaling to account for the spin dependence in the initial conditions (section \ref{subsec:runSim}), not other simulation parameters (section \ref{subsec:windPower}). Given the simultaneous wind and jet driving, we postulate that our simulations could represent a radio-loud AGN or an intermediate spectral state of an X-ray binary (section \ref{sub:simjetwind}). As the expanding amount of observational data and techniques increasingly enables probing jet geometries \citep[e.g.,][]{Zdziarski2022ApJ...925..189Z,Yan2026ApJ..1007L..49Y}, future observational work will be able to provide important comparisons to our simulations. Future work will explore the impact of resistivity on the launching of these outflows. It will also be interesting to compare and calibrate the resistivity profiles with extracted anomalous resistivities from turbulence-resolving disk simulations \citep[e.g.,][]{Jacquemin-Ide2025arXiv}.

\section*{Acknowledgements}
The authors would like to thank Chandra Singh for the valuable feedback. JS and ND are supported by Vici research program VI.C.242.101 (PI Degenaar) financed by the Dutch Research Council (NWO) under grant ID https://doi.org/10.61686/LIRQH96389. The authors used Claude to assist with derivations. All AI-assisted calculations were independently checked by the authors, who take full responsibility for the content of the manuscript.

\section*{Data Availability}
The data underlying this article will be shared on reasonable request to the corresponding author.



\bibliographystyle{mnras}
\bibliography{references} 




\appendix
\section{Spin dependence of the initial densities}\label{ap:densInit}
\begin{figure*}
    \centering
    \includegraphics[width=0.8\linewidth]{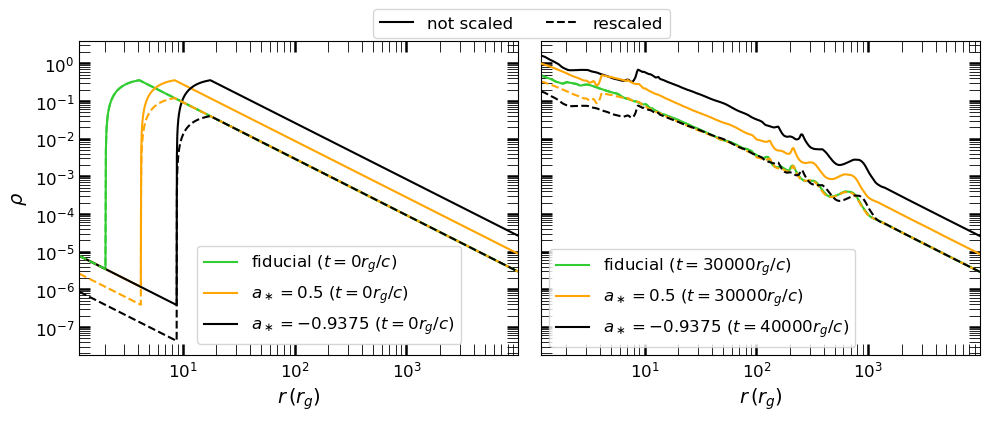}
    \caption{Left panel: Density profiles along the midplane at the initial timestep for the fiducial (green line), $a_*=0.5$ (orange lines), and the $a_*=-0.9375$ (black lines) simulations. The solid lines represent the unscaled densities, while the rescaled densities are plotted with dashed lines. Right panel: same as the left panel, but showing the densities at the respective last timesteps.}
    \label{fig:densityMidPlaneSlice}
\end{figure*}

The initial density profiles (eq.\ref{eq:rho}) depend on $a_*$, as visualized by the solid lines in Fig.\ref{fig:densityMidPlaneSlice}. The density profiles shown are extracted along the midplane at the initial time (left panel) and the corresponding last timesteps (right panel). The initial density profiles peak at the same value, but at larger radii, the densities of the $a_*=0.5$ (orange lines) and $a_*=-0.9375$ (black lines) simulations are consequently larger than those of the fiducial one (green line). We can remove the density differences at large radii by dividing by $\rescale$ (eq.\ref{eq:rescale}), which results in the dashed lines. These density differences are still visible at the respective last simulation timesteps (right panel, solid lines); after division by $\rescale$, the density profiles align again at large radii (dashed lines). 

While the right panel exemplifies the lasting impact of the initial conditions, it also shows the physical impact of $a_*$: the visible drops in $\rho$ at $r\approx4.2r_g$ for the $a_*=0.5$ simulation and at $r\approx8.8r_g$ for the $a_*=-0.9375$ simulation occur at the respective $r_{ms}$ (eq.\ref{eq:rms}). Because of the matter infall beginning at larger radii, $\absB{u^r}$ is greater for these two simulations than for the fiducial one as well. At $r\approx2r_g$ at the midplane and comparing the respective last timesteps, $\absB{u^r}$ is $\approx1.23$ times greater in the $a_*=0.5$ simulation and $\approx2.63$ times greater in the $a_*=-0.9375$ simulation than in the fiducial one.  

\section{Light cylinder}\label{ap:lightCyl}
An observer corotating with a magnetic field line has a velocity \citep[][]{Pugliese2018EPJC...78...69P}
\begin{equation}
    u^\alpha = \zeta \anglB{\partial_t +\Omega_F\partial_\phi},
\end{equation}  
where $\Omega_F$ is the field rotation frequency and $\zeta$ depends with
\begin{equation}
    \zeta^{-2} = -\mathcal{Z}\anglB{\Omega_F^2g_{\phi\phi}+2\Omega_F g_{t\phi} + g_{tt}}
\end{equation}
on the norm of $u^\alpha$, $g_{\alpha\beta} u^\alpha u^\beta = -\mathcal{Z}$.

At the light cylinder surface, the observer moves at the speed of light, so $g_{\alpha\beta} u^\alpha u^\beta =0$. For $\zeta$ and thus $u^\alpha$ to stay finite requires 
\begin{equation}
    \Omega_F^2g_{\phi\phi}+2\Omega_F g_{t\phi} + g_{tt} = 0.
\end{equation}
In the 3+1 decomposition, this condition becomes \citep{Ripperda2019ApJS}
\begin{equation}
    -\alpha^2 + \gamma_{ij}\anglB{\beta^i + \Omega_F\updown{\delta}{i}{\phi}}\anglB{\beta^j + \Omega_F \updown{\delta}{j}{\phi}}=0.
\end{equation}
The field rotation frequency is \citep[e.g.,][]{McKinney2006MNRAS.368.1561M}
\begin{equation}
    \Omega_F = \frac{F_{tr}}{F_{r\phi}} = \frac{F_{t\theta}}{F_{\theta\phi}},
\end{equation}
where the Maxwell tensor is \citep[][]{Ripperda2019ApJS}
\begin{equation}
    F_{\mu\nu} = n_\mu E_\nu - n_\nu E_\mu + \sqrt{-g} \eta_{\mu\nu\lambda\delta} n^\lambda B^\delta.
\end{equation}
With $A$ representing either $\phi$ or $r$, the equation for the field rotation frequency is
\begin{equation}
    \Omega_F = \frac{-\alpha E_A - \sqrt{\gamma}\beta^j\eta_{A j k}B^k}{\sqrt{\gamma}\eta_{A\phi k}B^k}.
\end{equation}
\section{Mass loading of the wind}\label{ap:massloadwind}
\begin{figure}
    \centering
    \includegraphics[width=0.9\linewidth]{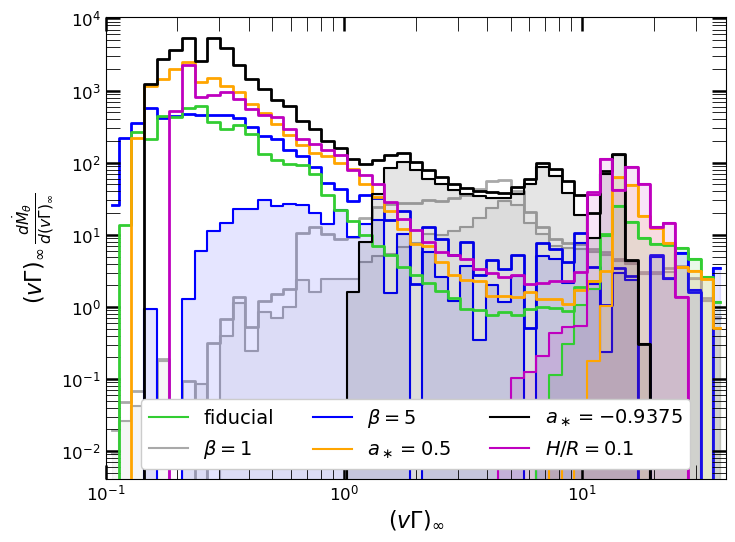}
    \caption{Distribution of $\mdot_\theta$ as a function of $\anglB{v\Gamma}_\infty$ for the wind region. All quantities are measured along an angle $4H/R$ as measured from the equatorial plane. The shaded regions show the portions of the distributions affected by the density floors (see also the discussion of Fig.\ref{fig:outFlowsVelocity}).}
    \label{fig:dmdGamma2}
\end{figure}
Fig. \ref{fig:dmdGamma2} shows the distribution of $\mdot_\theta$ as a function of $\anglB{v\Gamma}_\infty$ for the wind region along an angle of $4H/R$ as measured from the equatorial plane (i.e., the same slice as used for Fig. \ref{fig:outFlowsVelocity}, orange lines in Fig.\ref{fig:windWedgeRegions}). Here, the angular \mdotname $\mdot_\theta = 2\pi u^\theta \rho\sqrt{-g}$ is not spatially integrated. The shaded distributions are calculated for the regions affected by the density floors. 
For all simulations except the $\beta=1$ (grey) and $\beta=5$ (blue) ones, we define the floor-affected regions to be at smaller radii than the density minimum within 200 $r_g$ for every timestep (compare with Fig. \ref{fig:outFlowsVelocity}). As visible in the middle panel of Fig. \ref{fig:outFlowsVelocity}, the density profiles of the $\beta=1$ and $\beta=5$ simulations are more variable at larger radii, so we choose the first location larger than $10r_g$ at which the density profile is increasing as the boundary of the floor-affected region. 

Fig.\ref{fig:dmdGamma2} indicates a bimodal outflow distribution for all simulations except the $\beta=1$  and $\beta=5$ ones: a fraction of the outflow is launched at low velocities within the $\sigma < 1$ region, while another high-velocity wind exists where $\sigma>1$ (due to the density floors, we expect $\anglB{v\Gamma}_\infty$ to be even larger in the $\sigma>1$ region). While this bimodality persists when just plotting the $\anglB{v\Gamma}_\infty$ distribution (see also the bottom panel of Fig. \ref{fig:windWedgeRegions}), we are unable to gauge the effectiveness of the high-velocity wind in carrying away matter. As visualized by the shaded regions in Fig.\ref{fig:dmdGamma2}, bins at large $\anglB{v\Gamma}_\infty$ are affected by the density floors. Since the density profiles (middle panel of Fig. \ref{fig:outFlowsVelocity}) do not indicate any contributors to the density other than the floors, we infer that mass-loading is negligible for high-velocity winds (the lack of mass loading is discussed in section \ref{subsub:lackmassloading}).

\section{Force Balance}\label{ap:forceBalance}
To project forces along the poloidal magnetic field $B_p$, we introduce the normal vector 
\begin{equation}
    \hat{e}_P = \frac{B_P}{\absB{B_P}}.
\end{equation}
The poloidal component, is the total magnetic field minus its toroidal component $B_T$, \begin{equation}
    B^i_P = B^i - B^i_T.\label{eq:Bpoloidal}
\end{equation}
The toroidal component $B_T$ is parallel to $m^i$,
\begin{equation}
B^i_T = \frac{B^i m_i}{\gamma_{\phi\phi}} m^i = \frac{B_\phi}{\gamma_{\phi\phi}} m^i ,  \label{eq:Btoroidal} 
\end{equation}
where $m^i = \updown{\delta}{i}{\phi}$ is for an axisymmetric metric a Killing vector \citep{Misner1973grav.book.....M}, with a norm of $m^i$ of
\begin{equation}
    m^i m_i = g_{ji}m^i m^j = g_{ji}\updown{\delta}{\phi}{i}\updown{\delta}{\phi}{j}  = g_{\phi\phi}= \gamma_{\phi\phi}.
\end{equation}

\subsection{Electromagnetic forces}
The divergence of the electromagnetic stress-energy tensor is \citep{Baumgarte2003ApJ...585..921B},
\begin{equation}
    \nabla_\mu T^{\mu\nu}= - F^{\nu \lambda}\mathcal{J}_\lambda, \label{eq:emq1}
\end{equation}
with the Maxwell tensor $F^{\nu\lambda}$ \citep{Ripperda2019ApJS}
\begin{equation}
\begin{aligned}
    F^{\nu\lambda} &= n^{\nu} E^\lambda - n^{\lambda} E^\nu - \anglB{-g}^{-1/2}\eta^{\nu\lambda\alpha\beta}n_\alpha B_\beta\\
    F_{\nu\lambda} &= n_{\nu}E_{\lambda}-n_{\lambda}E_{\nu}+(-g)^{-1/2}\,\eta_{\nu\lambda\alpha\beta}\,n^{\alpha}B^{\beta}
\end{aligned}
\end{equation}
and the electric four-current $\mathcal{J}^\mu$
\begin{equation}
   \mathcal{J}^\mu = n^\mu q + J^\mu. \label{eq:fourcurrent}
\end{equation}
The vector $n^\mu = \anglB{1/\alpha,-\beta^i/\alpha}$ is the four velocity, with $\alpha$ being the lapse function and $\beta^i$ the shift-three vector, and 
\begin{equation}
    J^i = q v^i + \frac{\Gamma}{\eta}\squB{E^i + \gamma^{-1/2}\eta^{ijk}B_j B_k - \anglB{v_k E^k}v^i}
\end{equation}
is the spatial current density. Lowering an index in eq.\ref{eq:emq1} yields $\nabla_\mu \updown{T}{\mu}{\nu} = -F_{\nu\lambda} \mathcal{J}^\lambda$, which is minus the four-force density. The spatial components of the four-force density, i.e., $-\nabla_\mu \updown{T}{\mu}{i}$, are 
\begin{equation}
   F_{i\lambda} \mathcal{J}^\lambda  = \anglB{-n_{\lambda}E_{i}+(-g)^{-1/2}\,\eta_{i\lambda\alpha\beta}\,n^{\alpha}B^{\beta}}\mathcal{J}^\lambda,
\end{equation}
because $n_i = 0$ \citep{Ripperda2019ApJS}. The first component contains the charge density, $q = -n_\lambda\mathcal{J}^\lambda$. To simplify the second term, one can use eq.\ref{eq:fourcurrent} and recognize that the part with $n^\mu q$ is zero due to the asymmetry of $\eta_{i\lambda\alpha\beta}$. Furthermore, $\eta_{i j\alpha\beta}\,n^{\alpha} = \eta_{i j\beta}$ \citep{Baumgarte2003ApJ...585..921B}, and $B$ is purely spatial, so $B^t=0$. The spatial components of the four-force density therefore are 
\begin{equation}
    F_{i\lambda} \mathcal{J}^\lambda  = q E_i + (-g)^{-1/2}\,\eta_{i j k}\,B^{k}J^j = q E_i + \anglB{J\times B}_i,
\end{equation}
Which consist of the electric force and the force due to magnetic acceleration. 

The current and the magnetic field can be splitted into a poloidal and toroidal component (see eq. \ref{eq:Btoroidal} and \ref{eq:Bpoloidal}), which gives 
\begin{equation}\begin{aligned}
    \anglB{J\times B} &= \squB{\anglB{J_T + J_P}\times \anglB{B_T + B_P}} \\&= \anglB{J_T\times B_T} + \anglB{J_P\times B_T} +\anglB{J_T\times B_P}+\anglB{J_P\times B_P}\\
    &=\anglB{J_P\times B_T} +\anglB{J_T\times B_P}+\anglB{J_P\times B_P}
    \end{aligned}
\end{equation}
The term $\anglB{J_P\times B_T}$ accelerates the wind along the poloidal magnetic field and provides a z-pinch perpendicular to it \citep[e.g.,][]{Vlahakis2003ApJ...596.1080V,Jannaud2023A&A...669A.159J}. The second term, $\anglB{J_T\times B_P}$, collimates the wind and acts perpendicular to the magnetic field line \citep[see also, e.g.,][]{Vlahakis2003ApJ...596.1080V}. The last term $\anglB{J_P\times B_P}$ acts in the $\hat{\phi}$ direction \citep[e.g.,][]{Chen2021ApJ...906..105C}. Consequently, the electromagnetic force contributions of interest, projected along $B_p$, are 
\begin{align}
    \mathrm{electric} &= q E_i \hat{e}_P^i\\
    \mathrm{magnetic\,acceleration} &= \anglB{J_P\times B_T}_i \hat{e}_P^i.
\end{align}

\subsection{Non-electromagnetic forces}
The momentum equation is \citep{Ripperda2019ApJS}
\begin{equation}\begin{aligned}
    \partial_t \anglB{\gamma^{1/2}S_i}+&\partial_j\squB{\gamma^{1/2}\anglB{\alpha \updown{W}{j}{i}-\beta^j S_i}} \\&= \gamma^{1/2}\anglB{\frac{\alpha}{2}W^{j k }\partial_i \gamma_{jk}-U\partial_i \alpha+ S_k\partial_i \beta^k}.\label{eq:nonEm2}
\end{aligned}
\end{equation}

where $h$ is the enthalpy, $S_i$ the fluid three-momentum density 
\begin{equation}
    S_i = -\updown{\gamma}{\mu}{i} n^\alpha T_{\alpha\mu} = \alpha \updown{T}{t}{i} = \rho h \Gamma^2 v_i,
\end{equation} 
$U$ the conserved fluid energy density 
\begin{equation}
    U = T^{\mu\nu}n_\mu n_\nu = \alpha^2 T^{tt} = \rho h\Gamma^2 - p,
\end{equation}
and $W^{ij}$ the fluid stress-energy tensor 
\begin{equation}
    W^{ij} =\updown{\gamma}{i}{\mu}\updown{\gamma}{j}{\nu}T^{\mu\nu} =\rho h \Gamma^2 v^i v^j+p\gamma^{ij}.
\end{equation}

With Lie derivative with respect to $\beta^i$ \citep{Baumgarte2010nure.book.....B},
\begin{equation}\begin{aligned}
    \mathcal{L}_\beta S_j &= \beta^i\partial_i S_j + S_i \partial_j \beta^i\\ \mathcal{L}_\beta S^j &= \beta^i\partial_i S^j - S^i \partial_i\beta^j,
\end{aligned}
\end{equation}
the covariant derivative
\begin{equation}
    \begin{aligned}
        \covdev_i \beta^i &= \gamma^{-1/2}\partial_i\anglB{\gamma^{1/2}\beta^i}\\
        \covdev_i W^i_j &= \gamma^{-1/2}\partial_i\anglB{\gamma^{1/2}W^i_j} - \frac{1}{2}W^{ik}\partial_j \gamma_{ik},
    \end{aligned}
\end{equation}
and the trace of the extrinsic curvature,
\begin{equation}
    K = \alpha^{-1}\anglB{\covdev_i\beta^i - \partial_t \ln\gamma^{1/2}},
\end{equation}
eq.\ref{eq:nonEm2} can be recast as (assuming a stationary metric)
\begin{equation}\begin{aligned}
    0=&\gamma^{1/2}\anglB{\partial_t-\mathcal{L}_\beta}S_i-\gamma^{1/2}\anglB{S_i \covdev_j\beta^j}\\& + \gamma^{1/2}\anglB{\alpha \covdev_j W^j_i + W^j_i \covdev_j\alpha + U \covdev_i \alpha}.\label{eq:nonEm3}\end{aligned}
\end{equation}
Dividing by $\gamma^{1/2}$, and introducing the momentum vector $ w_i =h\Gamma v_i$ and the rest-mass density as measured in a Eulerian frame $D=\rho\Gamma$ \citep{Ripperda2019ApJS} allows writing eq.\ref{eq:nonEm3} as
\begin{equation}
    \begin{aligned}
       0 = &\anglB{\partial_t-\mathcal{L}_\beta}\anglB{Dw_j}-\squB{\anglB{Dw_j} \covdev_i\beta^i} \\&+ \alpha \covdev_i \anglB{Dv^iw_j} + \anglB{Dv^iw_j} \covdev_i\alpha+ \rho h \Gamma^2 \covdev_j \alpha +\alpha \covdev_j p\\
       0=&\partial_t\anglB{Dw_j}-\beta^i D \partial_i w_j-\beta^i w_j \partial_i D- Dw_i\partial_j \beta^i\\
       &-\anglB{Dw_j}\gamma^{-1/2}\partial_i\anglB{\gamma^{1/2}\beta^i}+\alpha w_j\covdev_i \anglB{D v^i } +\alpha D v^i\covdev_i w_j\\
       &+\anglB{Dv^iw_j} \covdev_i\alpha +\rho h \Gamma^2 \covdev_j \alpha +\alpha \covdev_j p.
    \end{aligned}
\end{equation}
Given the continuity equation \citep[][]{Ripperda2019ApJS}
\begin{equation}
    \partial_t D + \covdev_i\anglB{\alpha v^i D - \beta^i D} = 0,
\end{equation}
terms cancel, leaving after a division by $\alpha$
\begin{equation}
    0=D \squB{\alpha^{-1}\anglB{\partial_t- \mathcal{L}_\beta}+ v^i\covdev_i}w_j+\frac{\rho h \Gamma^2}{\alpha} \covdev_j \alpha + \covdev_j p.\label{eq:nonem4}
\end{equation}
Eq.\ref{eq:nonem4} consists of inertia and force terms. To isolate the forces that act along $B_p$, we split the velocity and the momentum vector in a poloidal and toroidal component,
\begin{align}
    v^i &= \frac{v_\phi}{\gamma_{\phi\phi}}m^i + v_P^i\label{eq:vi}\\
    w_i &=h\Gamma v_i =\frac{h\Gamma v_\phi}{\gamma_{\phi\phi}}m_i + h\Gamma v^P_i.\label{eq:wi}
\end{align}
Moving inertia terms to the left side of eq. \ref{eq:nonem4} and force terms to the right, yields the forces, projected along $B_p$,
\begin{align}
    \mathrm{pressure} &= -\partial_i p \hat{e}_P^i\\
    \mathrm{gravity} &= -\frac{\rho h \Gamma^2}{\alpha}\partial_i\alpha\hat{e}_P^i\\
    \mathrm{centrifugal} &= \frac{\rho h}{2} \anglB{\frac{\Gamma v_\phi}{\gamma_{\phi\phi}}}^2\partial_i\gamma_{\phi\phi}\hat{e}_P^i\\
    \mathrm{coriolis} &= -\frac{\rho h v_\phi\Gamma^2 v^{P,j}}{\gamma_{\phi\phi}}\anglB{\partial_j m_i - \partial_i m_j}\hat{e}_P^i\\
    \mathrm{shift/frame\,dragging} &= \frac{\rho \Gamma \mathcal{L}_\beta}{\alpha}w_i\hat{e}_P^i.
\end{align}
Note that because the Coriolis force and the contributions coming from the heating function are much smaller compared to the other forces, they are omitted in Fig. \ref{fig:forcesLines}.
\bsp	
\label{lastpage}
\end{document}